\documentclass[longauth]{aa} 

\usepackage[desactivate]{linenoaa}
\usepackage{graphicx}
\usepackage{txfonts}
\usepackage[colorlinks=true, linkcolor=blue, citecolor=blue, filecolor=blue, urlcolor=blue]{hyperref}
\usepackage{natbib}
\usepackage{xcolor}
\usepackage{multirow}
\usepackage[version=4]{mhchem}
\usepackage{placeins}
\usepackage{siunitx}
\usepackage{bm}
\usepackage{xspace}
\usepackage{array}

\graphicspath{{./}{figures/}}

\newcommand{\jwst}{JWST\xspace}                               
\newcommand{\hst}{HST\xspace} 
\newcommand{\spitzer}{\textit{Spitzer}\xspace}

\newcommand{\nirspec}{NIRSpec\xspace}  

\newcommand{\teatro}{\texttt{TEATRO}\xspace}
\newcommand{\cascade}{\texttt{CASCADe}\xspace}
\newcommand{\vulcan}{\texttt{VULCAN}\xspace}
\newcommand{\arcis}{\texttt{ARCiS}\xspace}
\newcommand{\taurex}{\texttt{TauREx}\xspace}
\newcommand{\petitradtrans}{\texttt{petitRADTRANS}\xspace}
\newcommand{\pyratbay}{\texttt{PYRAT\;BAY}\xspace}
\newcommand{\exotethys}{\texttt{ExoTETHyS}\xspace}

\newcommand{\mic}{\textmu m\xspace}
\newcommand{\eg}{{e.g.}\xspace}

\begin{document}

   \title{Detection of CO$_2$, CO, and H$_2$O in the atmosphere of the warm sub-Saturn HAT-P-12\,b}

   \titlerunning{CO$_2$, CO, and H$_2$O in the atmosphere of HAT-P-12\,b}
   \authorrunning{N. Crouzet et al.}

\author{
N. Crouzet\inst{1,2}
\and
B. Edwards\inst{3}
\and
T. Konings\inst{4}
\and
J. Bouwman\inst{5}
\and
M. Min\inst{3}
\and
P.-O. Lagage\inst{6}
\and
L. B. F. M. Waters\inst{7,3}
\and
J. P. Pye\inst{8}
\and \\
L. Heinke\inst{4,9,10}
\and
M. Guedel\inst{11,12}
\and
Th. Henning\inst{5}
\and
B. Vandenbussche\inst{4}
\and
O. Absil\inst{13}
\and
I. Argyriou\inst{4}
\and
D. Barrado\inst{14}
\and \\
A. Boccaletti\inst{15}
\and
C. Cossou\inst{16}
\and
A. Coulais\inst{17}
\and
L. Decin\inst{4}
\and
R. Gastaud\inst{16}
\and
A. Glasse\inst{18}
\and
A. M. Glauser\inst{12}
\and
F. Lahuis\inst{19}
\and \\
G. Olofsson\inst{20}
\and
P. Patapis\inst{12}
\and
D. Rouan\inst{15}
\and
P. Royer\inst{4}
\and
N. Whiteford\inst{21}
\and
L. Colina\inst{22}
\and
G. Östlin\inst{23}
\and
T. P. Ray\inst{24}
\and \\
E. F. van Dishoeck\inst{1}
}

\institute{
Leiden Observatory, Leiden University, P.O. Box 9513, 2300 RA Leiden, The Netherlands \\
\email{crouzet@strw.leidenuniv.nl}
\and
Kapteyn Astronomical Institute, University of Groningen, P.O. Box 800, 9700 AV Groningen, The Netherlands
\and
SRON Netherlands Institute for Space Research, Niels Bohrweg 4, 2333 CA Leiden, The Netherlands
\and
Institute of Astronomy, KU Leuven, Celestijnenlaan 200D, 3001 Leuven, Belgium
\and
Max-Planck-Institut für Astronomie (MPIA), Königstuhl 17, 69117 Heidelberg, Germany
\and
Université Paris-Saclay, Université Paris Cité, CEA, CNRS, AIM, F-91191 Gif-sur-Yvette, France
\and
Department of Astrophysics/IMAPP, Radboud University, PO Box 9010, 6500 GL Nijmegen, The Netherlands
\and
School of Physics \& Astronomy, Space Park Leicester, University of Leicester, 92 Corporation Road, Leicester, LE4 5SP, UK
\and
School of GeoSciences, University of Edinburgh, Edinburgh, EH9 3FF, UK
\and
Centre for Exoplanet Science, University of Edinburgh, Edinburgh, EH9 3FD, UK
\and
Department of Astrophysics, University of Vienna, Türkenschanzstr. 17, 1180 Vienna, Austria 
\and
ETH Zürich, Institute for Particle Physics and Astrophysics, Wolfgang-Pauli-Str. 27, 8093 Zürich, Switzerland
\and
STAR Institute, Université de Liège, Allée du Six Août 19c, 4000 Liège, Belgium
\and
Centro de Astrobiología (CAB), CSIC-INTA, ESAC Campus, Camino Bajo del Castillo s/n, 28692 Villanueva de la Cañada, Madrid, Spain
\and
LESIA, Observatoire de Paris, Université PSL, CNRS, Sorbonne Université, Université de Paris Cité, 5 place Jules Janssen, 92195 Meudon, France
\and
Université Paris-Saclay, CEA, IRFU, 91191, Gif-sur-Yvette, France
\and
LERMA, Observatoire de Paris, Université PSL, CNRS, Sorbonne Université, Paris, France 
\and
UK Astronomy Technology Centre, Royal Observatory Edinburgh, Blackford Hill, Edinburgh EH9 3HJ, UK
\and
SRON Netherlands Institute for Space Research, PO Box 800, 9700 AV, Groningen, The Netherlands
\and
Department of Astronomy, Stockholm University, AlbaNova University Center, 106 91 Stockholm, Sweden
\and
Department of Astrophysics, American Museum of Natural History, New York, NY 10024, USA
\and
Centro de Astrobiología (CAB, CSIC-INTA), Carretera de Ajalvir, 8850 Torrejón de Ardoz, Madrid, Spain
\and
Department of Astronomy, Oskar Klein Centre, Stockholm University, 106 91 Stockholm, Sweden
\and
School of Cosmic Physics, Dublin Institute for Advanced Studies, 31 Fitzwilliam Place, Dublin, D02 XF86, Ireland
}

   \date{Received ...; accepted ...}

   \abstract
   {The chemical composition of warm gas giant exoplanet atmospheres (with $T_{\rm eq} < 1000 \;\rm K$) is not well known due to the lack of observational constraints.}
   {HAT-P-12\,b is a warm, sub-Saturn-mass transiting exoplanet that is ideal for transmission spectroscopy. We aim to characterise its atmosphere and probe the presence of carbonaceous species using near-infrared observations.}
   {One transit of HAT-P-12\,b was observed in spectroscopy with \jwst NIRSpec in the 2.87--5.10~\mic range with a resolving power of $\sim$1000. The \jwst data are combined with archival observations from \hst WFC3 covering the 1.1--1.7~\mic range. The data were analysed using two data reduction pipelines and two atmospheric retrieval tools. Atmospheric simulations using chemical forward models were performed to interpret the spectra.}
   {CO$_2$, CO, and H$_2$O are detected at 12.2, 4.1, and 6.0 $\sigma$ confidence, respectively. Their volume mixing ratios are consistent with an atmosphere of $\sim$10$\times$ solar metallicity and production of \ce{CO2} by photochemistry. CH$_4$ is not detected and seems to be lacking, which could be due to a high intrinsic temperature with strong vertical mixing or other phenomena. SO$_2$ is also not detected and its production seems limited by low upper atmosphere temperatures ($\sim$500~K at $P\lesssim10^{-3}$~bar derived from one-dimensional retrievals), insufficient to produce it in detectable quantities ($\gtrsim$~800~K required according to photochemical models). H$_2$S is marginally detected using one data analysis method, but not by the other. Retrievals indicate the presence of clouds between 2 and 11~mbar using one data analysis method, and between 5 and 269~mbar using the other. The derived C/O ratio is below unity, but is not well constrained.}
   {This study points towards an atmosphere for HAT-P-12\,b that could be enriched in carbon and oxygen with respect to its host star, a possibly cold upper atmosphere that may explain the non-detection of \ce{SO2}, and a \ce{CH4} depletion that is yet to be fully understood. When including the production of \ce{CO2} via photochemistry, an atmospheric metallicity that is close to Saturn's can explain the observations. Metallicities inferred for other gas giant exoplanets based on their \ce{CO2} mixing ratios may need to account for its photochemical production pathways. This may impact studies on mass-metallicity trends and links between exoplanet atmospheres, interiors, and formation history.}
   
   \keywords{methods: observational -- techniques: spectroscopic -- eclipses -- planets and satellites: atmospheres -- planets and satellites: gaseous planets -- planets and satellites: individual: \mbox{HAT-P-12\,b}}

   \maketitle

\section{Introduction}
\label{sec:Introduction}

Carbon-bearing species are thought to be ubiquitous in the atmospheres of gas giant exoplanets; they are key contributors to chemical processes and provide clues to understanding their formation in protoplanetary disks \citep[\eg][]{Oberg2011, Fortney2013, Madhusudhan2014, Madhusudhan2019, Mordasini2016, Henning2024}. Before the launch of the \jwst, carbon monoxide had been detected in giant exoplanets using ground-based high-resolution spectroscopy \citep[\eg \mbox{HD~209458\,b}, \mbox{$\tau$~Bo\"otis\,b}, \mbox{HD~189733\,b}, \mbox{$\beta$~Pictoris\,b}, \mbox{TYC~8998-760-1\,b};][]{Snellen2010, Brogi2012, deKok2013, Snellen2014, Zhang2021} and integral field medium-resolution spectroscopy \citep[\mbox{HR~8799\,c},][]{Konopacky2013}. Methane had been detected in giant exoplanets using ground-based high-resolution spectroscopy \citep[\mbox{HD~102195\,b};][]{Guilluy2019} and spectral differential imaging \citep[\mbox{GJ~504\,b};][]{Janson2013}; weak methane absorption had been reported using integral field medium-resolution spectroscopy \citep[\mbox{HR~8799\,b};][]{Barman2011}; and possible hints of methane had been found in Neptune and sub-Neptune-sized exoplanets observed in photometry with \spitzer and low-resolution spectroscopy with the \textit{Hubble} Space Telescope (\hst) during their transits \citep[\mbox{GJ~436\,b}, \mbox{K2-18\,b};][]{Beaulieu2011, Bezard2022}. Constraining the presence and abundances of several carbon-bearing molecules for the same exoplanet remained challenging with these observations, and carbon dioxide remained out of reach.

A new era began with the \jwst. Carbon dioxide has been found in the atmospheres of close-in gas giant exoplanets such as \mbox{WASP-39\,b}, \mbox{HD~149026\,b}, \mbox{HD~209458\,b}, and \mbox{WASP-107\,b}; \mbox{WASP-39\,b} and \mbox{WASP-107\,b} also show CO \citep{JWST_Transit_ERS_2022, Rustamkulov2023, alderson_so2, Bean2023, Xue2024, Sing2024, Welbanks2024}. Methane has been detected in the warm gas giants \mbox{WASP-80\,b} and \mbox{WASP-107\,b}, and on the nightside of the ultra-hot gas giant \mbox{WASP-121\,b} \citep{bell_w80, Sing2024, Welbanks2024, Evans-Soma2025}. Methane depletion inferred from its low abundance or non-detection has been reported for gas giants including at atmospheric temperatures below 1000~K, which is puzzling since chemistry drives carbon into \ce{CH4} at these low temperatures \citep[\mbox{WASP-107\,b}, \mbox{WASP-39\,b}, \mbox{WASP-43\,b};][]{dyrek_w107, Sing2024, Welbanks2024, Ahrer2023, Bell2024}. A hot interior, low C/O ratio, disequilibrium chemistry, or the presence of clouds are possible explanations for this methane depletion \citep{dyrek_w107, Sing2024, Welbanks2024, Ahrer2023, Bell2024, Molaverdikhani2019b, Molaverdikhani2019a, Molaverdikhani2020}.
Both CO$_2$ and CH$_4$ have been identified in the atmospheres of several sub-Neptunes \citep[\mbox{K2-18\,b}, \mbox{TOI-270\,d}, \mbox{TOI-732\,c};][]{Madhusudhan2023, Benneke2024, Cabot2024}.
\jwst has also been used to link the chemical properties and structure of protoplanetary disks to the formation of exoplanets \citep{Henning2024}.

Photochemistry in exoplanet atmospheres has been studied theoretically and compared to observations \citep[\eg][]{Line2010, Moses2011, Tsai2021, Hobbs2021, polman_h2s_so2}. Evidence for photochemistry in an exoplanet atmosphere has been reported by \citet{tsai_so2} with the detection of a \ce{SO2} feature in the near-infrared spectrum of the hot Jupiter \mbox{WASP-39\,b} \citep{alderson_so2, Rustamkulov2023}. \citet{dyrek_w107} detected \ce{SO2} in the atmosphere of the warm Neptune \mbox{WASP-107\,b} in the mid-infrared with the \jwst Mid-Infrared Instrument \citep[MIRI,][]{Rieke2015, Wright2015} in low-resolution spectroscopy \citep[LRS,][]{kendrew_miri}, extending the range of atmospheric temperatures at which photochemistry has been found to occur and alter chemical equilibrium. The presence of \ce{SO2} was also confirmed for \mbox{WASP-39\,b} in the mid-infrared \citep{Powell2024} and for \mbox{WASP-107\,b} in the near-infrared \citep{Sing2024, Welbanks2024}. These detections show that photochemistry driven by ultraviolet (UV) radiation is key in the atmospheres of close-in gas giant exoplanets; \ce{SO2} is the only molecule detected to date that evidences such processes. The detection of hazes in gas giants may also provide hints for photochemistry, although condensation chemistry seems generally favoured as the general source of such hazes \citep[\eg][]{sing2016}.

\mbox{HAT-P-12\,b} \citep{Hartman2009} is a low-density, warm sub-Saturn-mass transiting exoplanet ($M_{\rm p} \approx 0.2\;\rm M_{Jup}$, $R_{\rm p} \approx 0.92\;\rm R_{Jup}$, $\rho_{\rm p} \approx 0.32 \rm \;g\,cm^{-3} $, $T_{\rm eq} \approx 955 \;\rm K$) in a 3.21-day period orbit around a K5 dwarf \citep[$T_{\rm eff} \approx 4665 \;\rm K$;][]{Mancini2018}. Its low density and large scale height combined with the brightness of its host star (K-magnitude of 10.1) make it an excellent object for transmission spectroscopy. 
A Rayleigh scattering slope due to clouds and hazes was detected with the \hst Space Telescope Imaging Spectrograph (STIS) by \citet{sing2016}. Observations with the \hst Wide Field Camera~3 (WFC3) in stare mode resulted in a non-detection of water vapour, which was attributed to a cloudy or hazy atmosphere \citep{Line2013, sing2016}. After implementation of WFC3 spatial scans \citep{McCullough2012}, a muted water vapour feature attenuated by clouds was detected at 1.4~\mic, the presence of photochemical hazes composed of small particles such as soot or tholins in the upper atmosphere was inferred from \hst STIS, and a metallicity higher than 60 times solar and a solar or subsolar C/O ratio were found after including \spitzer Infrared Array Camera (IRAC) photometric measurements at 3.6 and 4.5~\mic; the difference between the 3.6 and 4.5~\mic photometric fluxes points towards a possible CO$_2$ absorption feature \citep{wong_h12}. 
In contrast, ground-based spectrophotometric observations ruled out the presence of a Rayleigh slope extending over the entire optical wavelength range and suggested that grey absorption by a cloud layer could explain the flat transmission spectrum \citep{Mallonn2015}.
A reanalysis of the \hst STIS and ground-based observations revealed a much weaker Rayleigh slope than previously reported \citep{Alexoudi2018}.
Ground-based optical observations with the Large Binocular Telescope (LBT) also pointed to a relatively flat spectrum and found a supersolar metallicity, a C/O ratio close to solar, and confirmed the presence of high-altitude clouds \citep{Yan2020}.
Another study from ground-based observations with the Gran Telescopio Canarias (GTC) found evidence for stellar contamination that can explain the blueward slope observed with \hst STIS, no strong evidence for a cloudy or hazy atmosphere, and reported a high C/O ratio \citep{Jiang2021}.
Potassium and sodium have been tentatively detected with \hst STIS and with high-resolution spectroscopy at the Subaru telescope, respectively \citep{sing2016, Alexoudi2018, Deibert2019}.
\mbox{HAT-P-12\,b} has also been included in numerous population studies \citep[\eg][]{Barstow2017, Fisher2018, tsiaras_pop, Pinhas2018, Pinhas2019, edwards_hst_pop} and has been used in a general circulation model (GCM) study of a hot Jupiter sample \citep{Kataria2016}.
The orbit of \mbox{HAT-P-12\,b} is misaligned with respect to the rotation axis of its host star, with a projected spin–orbit angle of approximately $-54^{\circ}$ \citep{Mancini2018}.
The \mbox{HAT-P-12} system is similar to \mbox{WASP-107} \citep{Anderson2017}. \mbox{HAT-P-12\,b} and \mbox{WASP-107\,b} both have very low densities, equilibrium temperatures (i.e. the equivalent blackbody temperature of the planet if heated only by its host star) below 1000~K, and close-in misaligned orbits around a K5 or K6 star, and thus they provide an interesting comparison. \mbox{HAT-P-12\,b} is 1.7 times more massive than \mbox{WASP-107\,b}.

Discrepancies between different reductions and retrievals of the same observations have often been reported in transiting exoplanet spectroscopy \citep[\eg][]{w77_hst_jwst, Constantinou2023}. Assumptions and methods vary between analysis tools and the cause of these discrepancies is not always understood. To mitigate them and better assess the results, it has become common to use redundant tools in a single study, especially when using new instruments and exploring new spectral domains \citep[\eg][]{JWST_Transit_ERS_2022}.

\mbox{HAT-P-12\,b} is part of the sample of exoplanets and brown dwarfs observed by the ExoMIRI program of the \jwst MIRI European Consortium Guaranteed Time Observations. Three transits have been observed with the Near Infrared Imager and Slitless Spectrograph (NIRISS) single object slitless spectroscopy (SOSS) mode \citep{Doyon2012, Albert2023}, the Near Infrared Spectrograph (NIRSpec) with the G395M grating \citep{Birkmann2022, Jakobsen2022}, and MIRI LRS, covering a wavelength range from 0.6 to 12~\mic in total. This paper presents the analysis of the NIRSpec observation.  Here we use two data reduction pipelines and two retrieval codes, and perform retrievals on the spectra obtained from both reductions.
Section~\ref{sec:Observations} summarises the observation, Sect.~\ref{sec:Data reduction} describes the data reduction, and Sect.~\ref{sec:Atmospheric Modelling} defines the atmospheric modelling. The results are reported in Sect.~\ref{sec:Results} and are discussed in Sect.~\ref{sec:Discussion}. Our conclusions are given in Sect.~\ref{sec:Conclusion}.

\section{Observations}
\label{sec:Observations}

One transit of \mbox{HAT-P-12\,b} was observed with \jwst NIRSpec between 11 February 18:33 and 12 February 00:53, 2023 UTC (59986.7729 to 59987.0370 MJD) through the Guaranteed Time Observations of the MIRI European Consortium (GTO program 1281, PI: P.-O. Lagage). The observation was done in the bright object time-series (BOTS) mode with the medium-resolution grating G395M and filter F290LP, that covers the 2.87--5.10~\mic wavelength range with a resolving power of $\sim$1000. The medium-resolution gratings have the advantage of having all the light falling onto a single detector, NRS1. Target acquisition was done using the wide aperture target acquisition (WATA) method on the science target. The science observations used the 1.6" $\times$ 1.6" wide aperture S1600A1, the $2048 \times 32$ pixels subarray SUB2048, and the NRSRAPID readout pattern. The pixel size on the sky is 0.1~arcsec/pixel and the full spectral resolution is 0.0018~\mic/pixel. The observations were taken in one exposure of total duration 22834~s (6.34~h) for an effective exposure time of 22081~s (6.13~h). We used 30 groups per integration, one frame per group, and a total of 816 integrations. The effective integration time is 27.06~s per integration and the time between consecutive frames is 0.902~s. NIRSpec in-flight performance can be found in \citet{Espinoza2023} and \citet{Boker2023}.

\section{Data reduction}
\label{sec:Data reduction}

The \jwst data were reduced independently with two data reduction pipelines: the Transiting Exoplanet Atmosphere Tool for Reduction of Observations\footnote{\url{https://github.com/ncrouzet/TEATRO}} (\teatro) and \cascade\footnote{\url{https://gitlab.com/jbouwman/CASCADe}} \citep{bouwman_2023}. The procedures described below are standard for these pipelines and both have been benchmarked with other pipelines using \jwst MIRI data \citep{Bell2024, bouwman_2023, dyrek_w107}. Previous spectroscopic data from \hst were also reduced with \cascade and used in the analysis. The data described here may be obtained from \url{http://dx.doi.org/10.17909/58r8-2762}.

\subsection{\jwst NIRSpec reduction}

\subsubsection{\teatro reduction}
\label{sec:teatro reduction}

We processed the data using the \texttt{jwst} calibration software package version 1.10.2, Calibration Reference Data System (CRDS) version 11.16.19, and CRDS contexts 1097, 1100, and 1170 for stages 1, 2, and 3 respectively \citep{jwst2022, crds2022}. We started from the `uncal' files. For stage 1, we used the \texttt{Detector1} pipeline, set a jump rejection threshold of six, turned off the `flag\_4\_neighbors' parameter, and used the default values for all other parameters. For stage 2, we ran only the `AssignWcsStep', `Extract2dStep', `SourceTypeStep', and `WavecorrStep' steps of the \texttt{Spec2} spectroscopy pipeline. We did not use the `FlatFieldStep' in order to preserve suitable background regions around the two-dimensional (2D) spectrum. No photometric calibration was applied. To identify bad pixels, we used a running median along rows and searched for outliers. They were corrected for in each integration together with missing pixels by a rescaled bilinear interpolation, and the data quality (DQ) array was reset to 0 for all pixels.

In NIRSpec fixed-slit data, the 2D spectrum is tilted and bent along the dispersion direction, so a rectangular flux extraction aperture is not adequate. Instead, we computed the centre of the spectral trace in each column (cross-dispersion direction) by fitting a Gaussian function and finding the location of its maximum. These maxima were fitted with a third-order polynomial. This polynomial defines the centre of the aperture at each column and has the coefficients $18.3201$, $-1.0968 \times 10^{-2}$, $4.2911 \times 10^{-6}$, and $-1.2842 \times 10^{-10}$ from zeroth to third order, respectively. We computed the background for each column and each integration using regions above and below the aperture; their limits were defined using the same polynomial translated up or down and by the edges of the subarray. That method also mitigates the $1/f$-noise. We set the flux extraction aperture width to 5 pixels (0.5") with 2.5 pixels on each side from the centre, and the background region limits 2.5 pixels further away from the aperture. The corresponding polynomial coefficients were entered in the `extract1d' reference file, the background fit method was set to `median', and the flux was extracted using the `Extract1dStep' with `use\_source\_posn' and `apply\_apcorr' set to `False'. The white light flux was then computed with the `WhiteLightStep'.

For the spectroscopic analysis, we binned the spectra in 117 wavelength bins with central wavelengths from 2.85 to 5.17~\mic and a width of 0.02~\mic ($\sim$11~pixels). At native pixel resolution, adjacent spectral bins would be strongly correlated because the point spread function full width half maximum (FWHM) is about 2-pixel wide. In addition, this coarser spectral binning increases the signal-to-noise ratio in each bin and thus the precision of each data point in the light curves, which improves the light curve fits. We normalised each light curve by the out-of-transit stellar flux, measured on data points with orbital phases lower than $-0.016$ or greater than 0.016 (the mid-transit is at phase 0). For each light curve, we computed a running median filter using a 31-point window size, subtracted it, computed the standard deviation $\sigma$ of the residuals, and rejected points with residuals exceeding $3\,\sigma$; that rejection was iterated five times. This outlier rejection corrects for artefacts (such as cosmic rays) that were not well corrected in previous steps; we find generally fewer than ten outliers in each spectroscopic light curve. The same procedure was applied to the white light curve but only a few points deviated by more than $3\,\sigma$ and they were all within $4\,\sigma$ (which is expected if the distribution was purely Gaussian), so we kept all the points. These cleaned light curves were used for the transit fits. 

\subsubsection{\cascade reduction}
\label{subsubsec:Cascade_reduction}

We processed the data using the \texttt{jwst} calibration software package version 1.13.4 with the CRDS context 1188. We started from the `uncal' files and ran the `DQInitStep', `SaturationStep', `SuperBiasStep' ,`RefPixStep', and `LinearityStep' \texttt{Detector1} pipeline steps with the parameters and required calibration files defined by that CRDS context. To remove the observed $1/f$-noise \citep[see also][]{Rustamkulov_2022, Espinoza2023}, we opted to insert a correction in both the \texttt{Detector1} and the \texttt{Spec2} pipelines. First, we used the median value in time of all pairwise differences of the linearised detector ramps to create an estimate of the median spectroscopic signal. Next, using the \texttt{CASCADe-jitter}\footnote{\url{https://gitlab.com/jbouwman/CASCADe-jitter}} package version 0.9.5, we determined the spectral trace from the median image, and defined the source region as a 21-pixel-width aperture centred on the fitted trace position, starting from detector row 600. After that, we used all detector pixels outside of the source region to determine first a median background signal per detector column (cross-dispersion direction) and, after subtraction, a median background along the detector rows. We did this for each detector frame and subtracted that background from the linearised data product to correct for the $1/f$-noise. Finally, we applied the `JumpStep' and `RampFitStep' pipeline steps to produce the `rateints' data files.

For stage 2, we ran only the `AssignWcsStep', `Extract2dStep', and `SourceTypeStep' steps of the \texttt{Spec2} pipeline; no photometric or flatfield calibration was applied. In the `Extract2dStep', we used a slightly modified wavelength range of 2.5 to 5.27~\mic to define a slightly larger region of interest compared to the default values. Then, using the \texttt{CASCADe-filtering}\footnote{\url{https://gitlab.com/jbouwman/CASCADe-filtering}} package version 1.0.2, we identified any bad pixels or cosmic ray hits not identified by the \texttt{Detector1} pipeline, and cleaned all pixels flagged as `do not use'. We determined again the spectral trace for each integration from these cleaned images with the \texttt{CASCADe-jitter} package. The time-averaged polynomial coefficients of the spectral trace are $20.7340$, $-1.2729 \times 10^{-2}$, $4.3257 \times 10^{-6}$, and $-1.1599 \times 10^{-10}$ from zeroth to third order, respectively. Employing an identical procedure as used during the \texttt{Detector1} stage, we subtracted again a median background per detector row and column, per integration, to remove any remaining $1/f$-noise and the general infrared background emission.
Finally, we extracted the one-dimensional (1D) spectral time series from the spectral images using the `Extract1dStep' pipeline step. In this step, we used a 6-pixel-width extraction aperture centred on the wavelength-dependent source position defined by the polynomial above.

We checked the wavelength solution of our spectral data by cross-correlating the observed stellar spectrum against a stellar model of \mbox{HAT-P-12}, using a custom spectral response function. The latter was derived from NIRSpec observations of the A5IIIm star 2MASS J17430448+6655015 observed by the \jwst calibration program 1536 (PI: K. Gordon) and corresponding stellar model \cite[see][for further details on this program]{Gordon_2022}. We used the exact same calibration and spectral extraction procedures as for our \mbox{HAT-P-12} observations. We find no evidence for any wavelength shift in the observations of \mbox{HAT-P-12} to a precision well below a tenth of a resolution element.

\subsection{\jwst NIRSpec light curve fits}

\subsubsection{\teatro light curve fits}
\label{sec:teatro light curve fits}

We fitted the light curves obtained in Sect.~\ref{sec:teatro reduction} with a transit model and an instrument systematics model. The transit model was a quadratically limb-darkened transit light curve computed with the \texttt{exoplanet} package \citep{exoplanet:joss, exoplanet:zenodo}. We used a circular orbit \citep{Bonomo2017} with orbital period and mid-transit time from \citet{Kokori2022}, and a stellar mass, stellar radius, planet radius, inclination, and semi-major axis from \citet{Mancini2018}. We computed the limb-darkening coefficients with the \exotethys package \citep{Morello2020b, Morello2020} using a quadratic law, the STAGGER 2018 stellar atmosphere model grid \citep{Chiavassa2018}, and a stellar effective temperature, log$g$, and metallicity from \citet{Mancini2018} as listed in Table~\ref{tab:system parameters}. The instrument systematics model was a linear trend. Various systematics models have been used for previous exoplanet observations with \nirspec \citep{Espinoza2023, JWST_Transit_ERS_2022, Rustamkulov2023, alderson_so2}. Here using a higher order polynomial or an exponential did not improve the fits. The data quality and instrument stability were excellent and the light curves are nearly devoid of systematics (Sect.~\ref{sec:jwst nirspec light curves}). We fitted that model to the data using a Markov chain Monte Carlo (MCMC) procedure based on the \texttt{PyMC3} package \citep{Salvatier2016} and gradient-based inference methods as implemented in the \texttt{exoplanet} package.

First, we fitted the white light curve with the mid-transit time, planet-to-star radius ratio, impact parameter, limb-darkening coefficients, and the two coefficients of the linear trend as free parameters. We used wide priors centred on the values given in the references above and to those obtained from an initial least-squares fit to the out-of-transit data points for the linear trend. The orbital period, stellar mass, and stellar radius were fixed. Then, we fitted the spectroscopic light curves by fixing the mid-transit time and impact parameter to the values derived from the white light curve fit, and the wavelength-dependent limb-darkening coefficients to the values obtained with \exotethys. Only the planet-to-star radius ratio and the linear trend coefficients were allowed to vary. We set normal priors for the linear trend coefficients centred on the values obtained from an initial least-squares fit, a uniform prior for the radius ratio centred on the value derived from the white light curve fit, and allowed for wide search ranges. 

We ran two MCMC chains with $10\,000$ tune steps and $100\,000$ draws. Convergence was obtained for all parameters. We merged the posterior distributions of both chains and used their median and standard deviation to obtain final values for the parameters and uncertainties. We also verified that the values obtained from both chains were consistent.
We computed the transit depth uncertainty in two ways: from the standard deviation of its posterior distribution (i.e. the square of the radius ratio distribution) and by summing quadratically the standard deviations of the residuals of the in- and out-of-transit parts of the lightcurve divided respectively by the square root of their number of points. We kept the maximum of the two.

\subsubsection{\cascade light curve fits}
\label{subsubsec:Cascade_light curve}

We fitted the spectral light curves obtained in Sect.~\ref{subsubsec:Cascade_reduction} using the \cascade package \citep[see][for the details of the \cascade methodology and application]{carone_2021, bouwman_2023, dyrek_w107}. 
Before fitting, we binned the spectra to the same wavelength grid as \teatro. We also extracted spectra at full resolution and with a 0.0099~\mic bin width; the final exoplanet spectra are consistent with the adopted 0.02~\mic bin width.
In the \cascade analysis, we left out the first five integrations of the time series as they show a band-averaged response drift of about $0.06\,\%$, which is about four times larger than the maximal response drift seen in the rest of the light curve.
For the systematics model \citep[see][for details]{carone_2021}, we used as additional regression parameters the observing time (as a second order polynomial) and the median FWHM and median row position of the spectral trace (as linear functions).
The limb-darkening coefficients were calculated for each spectral channel in the same way as in Sect.~\ref{sec:teatro light curve fits}. We used the mid-transit time from the \teatro white light curve fit, adopted a circular orbit with a semi-major axis of 0.037~au, and an orbital inclination of $88.9^{\circ}$. These values are within $1\,\sigma$ of those reported by \citet{Mancini2018}. The uncertainties on the transit spectrum and band-averaged depth were determined by performing a bootstrap analysis.

\subsection{\hst WFC3 data reduction and light curve fits}

Two transits of \mbox{HAT-P-12\,b} were observed on 12 December 2015 and 31 August 2016 with the WFC3 instrument on board \hst using the 1.41~\mic grism (G141). The data were obtained as part of the General Observer program 14260 (PI: D. Deming). We refer to \citet{tsiaras_pop} and \citet{Panek2023} for details on the observations and the initial data analysis. We performed an independent calibration and light curve fitting of the \hst data using the \cascade package. Regarding details on the use of \cascade on \hst data, see \cite{carone_2021}. We ran \cascade using the same orbital and stellar parameters as used for the analysis of the \jwst NIRSpec light curves (Table~\ref{tab:system parameters}), except for the ephemeris. We used the mid-transit time values 2457368.783199 and 2457632.254063~BJD, respectively. After subtracting the best-fit light curve model, these mid-transit times resulted in residuals consistent with white noise.

Before fitting the spectral light curves, we binned the original spectral resolution of the \hst WFC3 data to a uniform wavelength grid with a bin width of 0.0095~\mic, to have independent bins and increase the signal-to-noise ratio of the light curves. For the first \hst orbit, the first five spatial scans were not used in our analysis as they show a response drift two to three times stronger than the maximal response drifts seen during the other orbits. For the systematics model \citep[see][for details]{carone_2021}, the additional regression parameters were the time variable and the trace position.
The uncertainties on the transit spectrum and band-averaged depth were estimated by performing a bootstrap analysis. For the retrieval analysis, we binned the spectrum to a slightly lower resolution, with a bin width of about 0.02~\mic, to further increase the signal-to-noise ratio per spectral channel and ensure that each spectral bin was independent.

\section{Atmospheric modelling}
\label{sec:Atmospheric Modelling}

\subsection{Retrieval models}
\label{sec:ret}

We fitted the \jwst NIRSpec and \hst WFC3 spectra using the ARtful modelling Code for exoplanet Science \citep[\arcis,][]{min_arcis}. We assumed that \mbox{HAT-P-12\,b} possesses a primordial atmosphere with a solar helium-to-hydrogen ratio (He/H$_2$ = 0.1765). The atmosphere was modelled between $P \in [10^{-10}, 10^2]$~bar using 100 plane-parallel layers uniformly partitioned in log-space. For all our retrievals, we employed an isothermal pressure-temperature ($P-T$) profile, setting bounds of $T \in [100, 1500]$~K. The temperature in the retrieval is constrained from its influence on the scale height (and thus on the amplitude of the transit spectral features) and the spectral shape of the molecular bands. We also looked at non-isothermal $P-T$ profiles but they did not change the inferences about the atmosphere. We assumed a simple grey-cloud model, where the only fitting parameter was the cloud pressure (CP) that had bounds of CP $\in [10^{-10}, 10^2]$~bar. 
For molecular absorption, we included H$_2$O \citep{polyansky_h2o}, CH$_4$ \citep{yurchenko_ch4}, CO \citep{li_co}, CO$_2$ \citep{yurchenko_co2}, H$_2$S \citep{azzam_h2s} and SO$_2$ \citep{underwood_so2}. 
\ce{NH3} and \ce{PH3} can also have contributions in this wavelength range but they were not detected in the first retrievals we performed (Bayes factors below 0.3 with both the \cascade and \teatro reductions) and the spectra can be fitted without them, so they were not included in further retrievals.
We also included collision-induced absorption from H$_2$-H$_2$ \citep{abel_h2-h2, fletcher_h2-h2} and H$_2$-He \citep{abel_h2-he} as well as Rayleigh scattering for all molecules. For each molecule, we set bounds on the volume mixing ratio (VMR) of $\rm log_{10}(VMR) \in [-15, 0]$, assuming a constant abundance with altitude. We used correlated k-tables to compute the opacities and all opacities were taken from the ExoMol database\footnote{\url{https://www.exomol.com/data/molecules/}} \citep{chubb_database}.
The star and planet parameters (stellar radius, planet radius, planet mass) were those from \citet{Mancini2018}.
We explored the parameter space using the nested sampling algorithm \texttt{MultiNest} \citep{Feroz_multinest,buchner_multinest} with 2500 live points and an evidence tolerance of 0.5, which is the value recommended in the documentation.\footnote{\url{https://github.com/JohannesBuchner/MultiNest/blob/master/README}} To compute the detection significance of each molecule, we ran additional retrievals in which each molecule was iteratively removed, computed the preference for their presence using the difference in the Bayesian evidence, and converted the Bayes factors into sigma significance using the formalism presented in \citet{Benneke2013}. 

We performed atmospheric retrievals on both the \teatro and \cascade spectra. These were conducted with the NIRSpec G395M data alone and with the addition of the \hst WFC3 G141 data to allow us to compare the inferences from these different datasets and reductions. As recovering absolute transit depths is difficult, particularly with \hst WFC3 \citep[\eg][]{edwards_hst_pop}, we mitigated for this by fitting for an offset \citep[\eg][]{yip_w96}. In our retrievals, we allowed the \hst WFC3 spectrum to be shifted relative to the \jwst NIRSpec spectrum due to the stronger systematics seen in that data. For each forward model sampled, \arcis finds the \hst offset that minimises the reduced chi-squared of the fit using a linear least-squares procedure.

As an additional comparison, we also performed a retrieval using the Tau Retrieval for Exoplanets (\taurex) 3.1 \citep{al-refaie_taurex3} on the NIRSpec G395M (\teatro) and \hst WFC3 G141 spectra. In this case, we used cross-sections instead of correlated k-tables, but these were also obtained from the ExoMol database. In \taurex, the offset $\alpha$ was a free parameter. The bounds for $\alpha$ were set to be extremely broad, with $\alpha \in [-500, +500]$~ppm. Other assumptions and priors were identical to the retrievals using \arcis.

\subsection{Forward models}
\label{subsec: forward models methods}

We investigated the physical and chemical processes that could lead to the retrieved volume mixing ratios by computing a set of forward models.
Since the forward models including photochemistry are too computationally expensive to use in a full statistical retrieval framework, we adopted an approximate forward modelling approach to arrive at estimates of the planet parameters that are consistent with the data and performed a moderate parameter study around these values. For the estimate of the parameters, we used a forward modelling approach that does not include photochemistry. We used the chemical relaxation method implemented in \arcis \citep{Kawashima2021} in combination with computations of the radiative-convective equilibrium \citep[see Appendix B of][for benchmarks of the \arcis radiative-convective solver]{Chubb2022}. This allows self-consistent iteration of the $P-T$ structure of the atmosphere using only the irradiation at the top of the atmosphere and the intrinsic temperature $T_\mathrm{int}$ \citep[i.e. the temperature that corresponds to the flux from the planet's interior; see \eg][]{Guillot2002, Guillot2010, Fortney2018} as input parameters. For the irradiation at the top of the atmosphere, we assumed that the radiation from the central star is isotropic at the top and is a scaled version of the stellar irradiation at the substellar point (with scaling parameter $\beta$). The value of $\beta$ at the limb of the planet is expected to range from 0 for the case of no heat redistribution to 0.25 for full heat redistribution from dayside to nightside. Next, we used a simple chi-square optimiser to find the optimal parameter set for the combined \hst and \nirspec (\teatro) spectra. Because the molecular abundances and the $P-T$ structure are computed from physics and chemistry, the resulting set of parameters is limited: planet radius, $\beta$, intrinsic temperature, C/O ratio, metallicity, and the eddy diffusion coefficient $K_{\rm zz}$. This last parameter determines the vertical mixing of molecules in the chemical relaxation scheme. Note that this modelling scheme does not include photochemical production or destruction of any molecules. Also, the optimisation procedure does not produce uncertainties, only the optimal values for the parameters. Therefore, the resulting best-fit values have to be interpreted with caution. 
It should also be noted that the models we used are 1D, which may lead to underestimating the atmospheric terminator temperature derived from transmission spectra (Sect.~\ref{sec:Retrieval outcomes}).

Then, we performed chemical kinetics simulations that include photochemistry using the open-source chemistry code \vulcan \citep{VULCANtsai2017}. 
We used a chemical network ({\small \uppercase{sncho\_photo\_network}}) that includes 89 species composed of H, C, O, N, and S, coupled by 1030 forward and backward thermochemical reactions.
Vertical mixing was included by the diffusion equation with accompanying eddy diffusion coefficient $K_{\rm zz}$; we used a constant $K_{\rm zz}$-profile for all models.
We used the pressure-temperature profile and $K_{\rm zz}$ value obtained from the optimisation procedure described above as inputs.
Photochemistry was included by adopting 60 photodissociation reactions for the main UV-absorbing species in \ce{H2}-dominated atmospheres.
We used the spectral energy distribution of \mbox{HAT-P-12} as constructed by the MUSCLES survey extension \citep{Behr2023MUSCLES}.
To initialise \vulcan, we used 150 log-spaced vertical layers between $10$ and $10^{-7}$~bar to ensure convergence throughout the atmosphere.
The elemental mixture was adjusted based on the input metallicity and carbon-to-oxygen ratio, and molecular abundances were computed from chemical equilibrium before the simulation started.
We did not include processes such as condensation, ion chemistry, or atmospheric escape.

We created synthetic spectra with \petitradtrans \citep{petitRADTRANSmolliere2019} using this $P-T$ structure and the abundances obtained from \vulcan, with the addition of a simple grey cloud deck.
We included line absorption opacities of \ce{SO2}, SO, \ce{H2S}, \ce{C2H2}, \ce{CO2}, CO, OH, HCN, \ce{NH3}, \ce{CH4}, and \ce{H2O}, Rayleigh scattering opacity of \ce{H2} and He, and continuum opacity from collision-induced absorption by \ce{H2-H2} and \ce{H2-He}.
In these models, we adopted a host star radius of 0.7\;R$_\odot$, a planet radius of 0.95\;R$_{\rm{Jup}}$, and a planet mass of 0.21\;M$_{\rm{Jup}}$.
Both \arcis and \petitradtrans have been benchmarked with a third code, \pyratbay, and produce similar outputs \citep{Cubillos2021a, Cubillos2021, Barstow2022}.

\section{Results}
\label{sec:Results}

\subsection{\jwst \nirspec light curves}
\label{sec:jwst nirspec light curves}

\begin{figure}
    \includegraphics[width=\columnwidth]{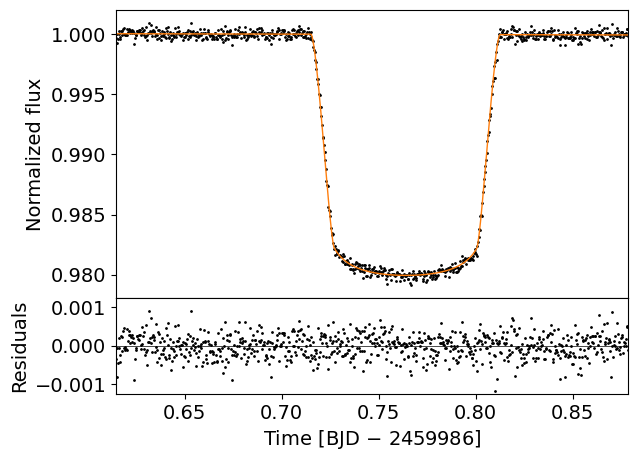}
    \caption{Transit white light curve of HAT-P-12\,b obtained with \jwst NIRSpec G395M (black points) and best-fit model (orange line). The residuals are shown in the bottom panel.}
    \label{fig:white light curve}
\end{figure}

\begin{figure}
    \includegraphics[width=0.95\columnwidth]{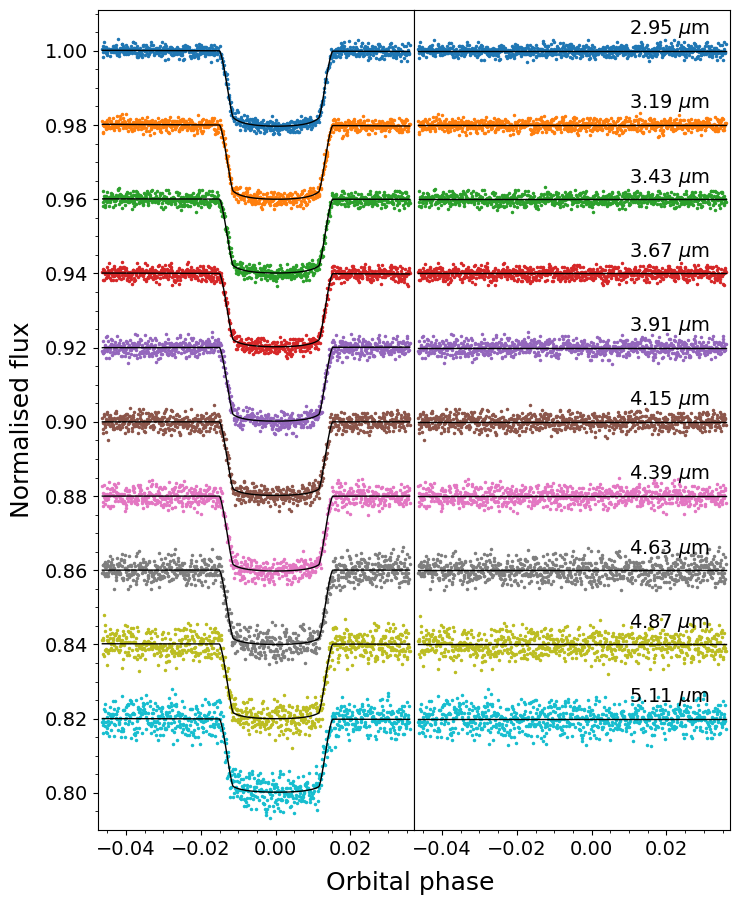}
    \caption{Sample of ten light curves out of 117 distributed over the $2.85-5.17 \; \mu$m wavelength range. Each light curve covers a $0.02 \; \mu$m wavelength bin. Left: Transit light curves (coloured points) and best-fit models (black lines). Right: Residuals. The central wavelength of each bin is indicated on the right. Systematic trends have not been removed in the left panel and are extremely small. These light curves were obtained with the \teatro reduction. The light curves are offset for clarity.}
    \label{fig:spectroscopic light curves}
\end{figure}

The white light curve obtained with \teatro is shown in Fig.~\ref{fig:white light curve}. The cadence is 28~s and the standard deviation $\sigma$ computed from the out-of-transit points before the fit is 306~ppm. Systematics are extremely small: the linear trend is only $-16$~ppm per hour. The variations of the position of the spectral trace in the spatial direction and of the FWHM have both a standard deviation of 0.003 pixels, which shows the extreme stability of the instrument. After fitting for the transit and systematics, the standard deviation of the residuals is 301~ppm. The system parameters derived from the white light curve fits are reported in Table~\ref{tab:system parameters}. 
The differences between those derived from \teatro and from \cascade are due to the fact that $a/R_{\star}$ is fixed for \teatro but is free for \cascade. When $a/R_{\star}$ is free for \teatro, the derived parameters are fully consistent with those from \cascade. The band-averaged transit depth derived with \cascade is $18850 \pm 16$~ppm.
The spectroscopic light curves also have very small systematics: with \teatro, these are well modelled by a linear trend with a slope that is between $-80$ and $+50$~ppm per hour depending on the spectroscopic channel. After fitting, the standard deviation of the residuals varies from 1000 to 3000~ppm from the shortest- to the longest-wavelength channels. A sample of these light curves is shown in Fig.~\ref{fig:spectroscopic light curves}.

\subsection{\jwst \nirspec transmission spectrum}

\begin{figure}
    \includegraphics[width=\columnwidth]{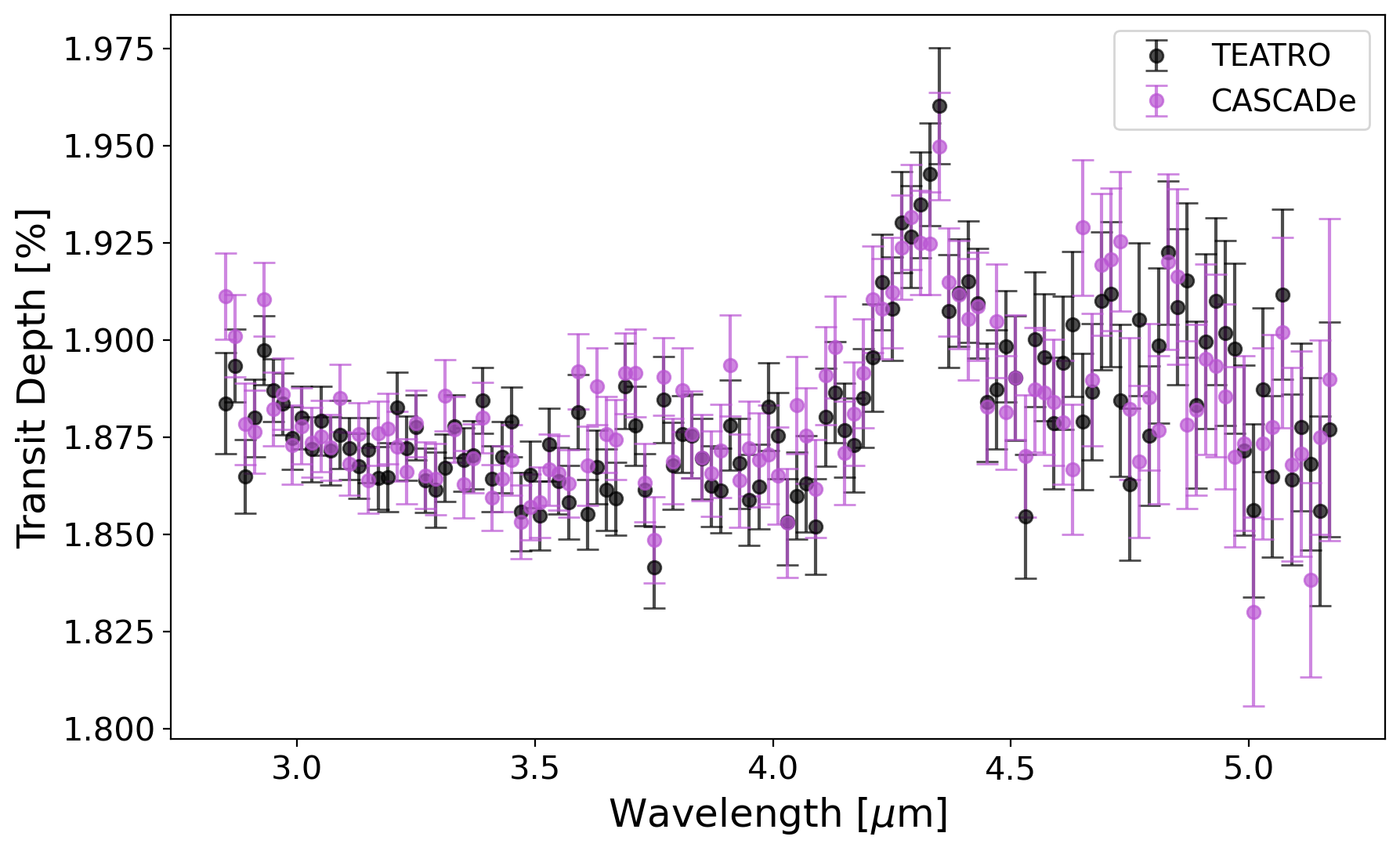}
    \caption{Transmission spectrum of HAT-P-12\,b obtained with \jwst NIRSpec G395M from our reductions with \teatro (black) and \cascade (purple). No vertical offset has been applied to the spectra.}
    \label{fig:spectrum}
\end{figure}

\begin{figure}
    \includegraphics[width=\columnwidth]{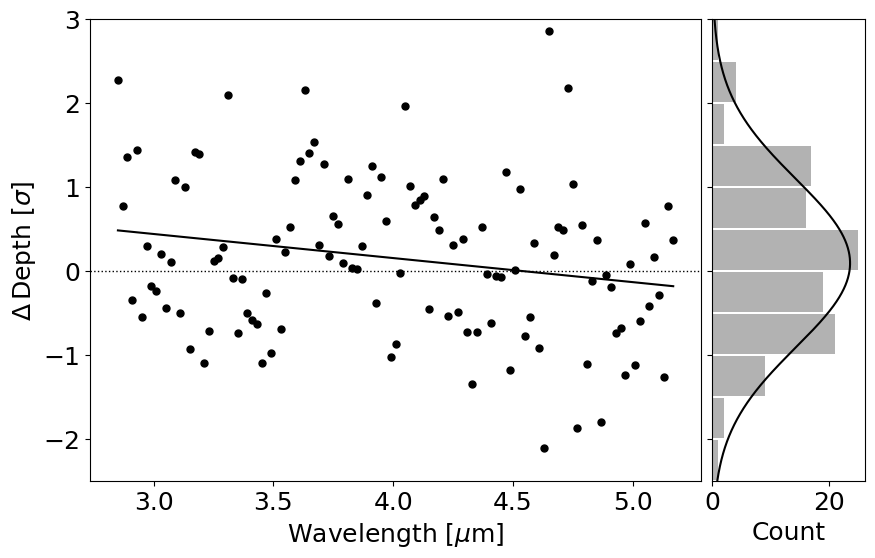}
    \caption{Left panel: Transit depth difference between the \cascade and \teatro data reductions expressed in number of standard deviations $\sigma$ as computed from Eq.~(\ref{eq:spectrum difference}), and best linear fit. Right panel: Histogram of that distribution, and best Gaussian fit.}
    \label{fig:spectrum diff}
\end{figure}

The two transmission spectra obtained from our independent reductions and light curve fits are shown in Fig.~\ref{fig:spectrum} and given in Table~\ref{tab:spectrum}. Visually, the spectra are highly similar. We computed their difference at each wavelength in terms of the number of standard deviations as 
\begin{equation}
\label{eq:spectrum difference}
    \Delta \delta(\lambda) = \frac{\delta_2(\lambda) - \delta_1(\lambda)}{\sqrt{\,\frac{1}{2}\,\left(\sigma_1^2(\lambda) + \sigma_2^2(\lambda)\right)}},
\end{equation}
where $\delta$ is the transit depth, $\lambda$ is the wavelength, and indices~1 and~2 refer to \teatro and \cascade, respectively. The denominator is the quadratic mean of the $1\,\sigma$ uncertainties of the data points of both spectra at a given wavelength; its mean value over all wavelengths is 137~ppm and the mean values of $\sigma_1$ and $\sigma_2$ are 133 and 139~ppm respectively.
These differences and their distribution are shown in Fig.~\ref{fig:spectrum diff}. We find that 69\% of the points agree within $1\,\sigma$, 95\% within $2\,\sigma$, and they all agree within $3\,\sigma$, which is expected if they would be normally distributed.
We fitted this distribution with a Gaussian using non-linear least-squares: the best fit has a mean of $0.09 \pm 0.11$ and a standard deviation of $1.00 \pm 0.11$, which indicates that both reductions are consistent within their uncertainties, ignoring any correlation with wavelength.
We also investigated a possible trend by performing a linear regression of $\Delta \delta(\lambda)$ with wavelength: the slope is $-0.29 \pm 0.13 \; \rm \mu m^{-1}$, the Pearson correlation coefficient is $-0.21$, and the p-value for a hypothesis test whose null hypothesis is that the slope is zero is 0.024. Thus there is a weak linear correlation with wavelength with an amplitude of $-0.66$ (about $-91\;\rm ppm$) over the full wavelength range, which remains below the $1\,\sigma$ level. The linear trend computed directly from the difference between both spectra $\delta_2(\lambda) - \delta_1(\lambda)$ gives similar results, with a slope of $-41 \pm 18 \; \rm ppm \; \mu m^{-1}$ and an amplitude of $-95\;\rm ppm$ over the full wavelength range. 
Notably, 19 consecutive \cascade points are higher than \teatro in the $3.55-3.9\;\mu$m range at the location of a potential \ce{H2S} feature, including five consecutive points higher by 100 to 200~ppm (one to two~$\sigma$) in the $3.6-3.7\;\mu$m range. 
The cause of these differences has not been identified but may be related to the parameters used or to the treatment of systematics in the light curves: \cascade uses a common mode analysis based on information available at all wavelengths whereas \teatro removes trends only on individual light curves. These differences have an impact on the inference on the species, in particular on the presence of \ce{H2S}.

\subsection{\hst WFC3 transmission spectrum}

\begin{figure}
   \begin{center}
    \includegraphics[width=0.98\columnwidth]{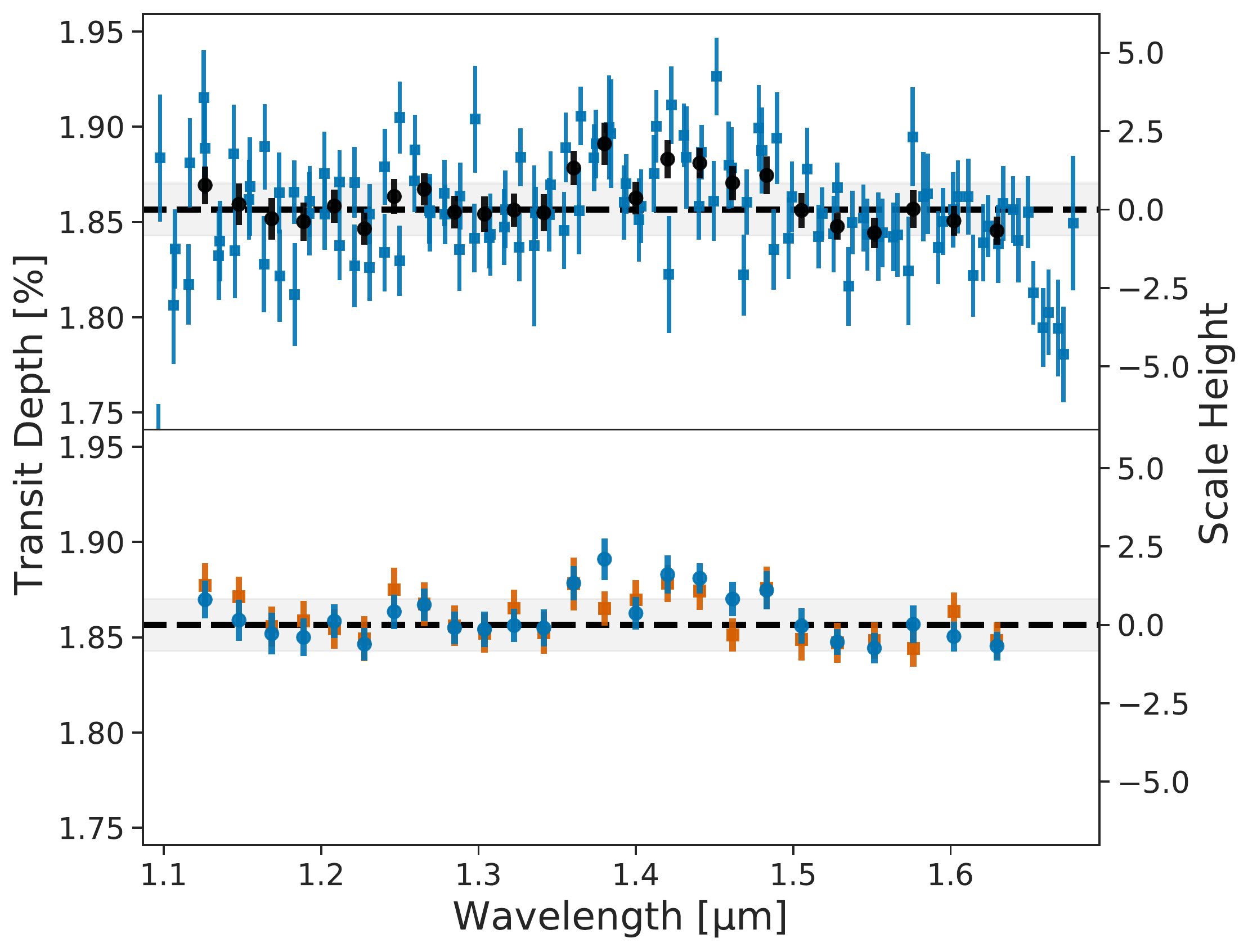}
 \end{center}
    \caption{Transmission spectrum of HAT-P-12\,b obtained with \hst WFC3 G141 using our \cascade reduction. The top panel shows the spectra for the two observed transits (blue squares) and the combined spectrum binned to a uniform spectral resolution of 0.02~\mic (black dots). The bottom panel shows a comparison of the spectrum derived in this study (blue dots) and the one derived by \cite{tsiaras_pop} (orange squares). In both panels, the band-averaged transit depth is indicated by the dashed horizontal line, and the shaded area represents its 95\% confidence interval. The spectrum from \cite{tsiaras_pop} has been shifted downwards by 150~ppm to the same mean transit depth as found by \cascade, for better comparison. The y-axis on the right is in units of planetary atmospheric scale height assuming a hydrogen-dominated atmosphere and an equilibrium temperature of $\sim$500~K as derived from the \arcis retrieval.}
    \label{fig:HST}
\end{figure}

The transit spectrum obtained from \hst WFC3 G141 is shown in Fig.~\ref{fig:HST}. We find a band-averaged transit depth of $18565 \pm 69$~ppm, which is about $4\,\sigma$ lower than the band-averaged transit depth obtained from \jwst NIRSpec using \cascade.
A comparison of the \hst spectrum derived using \cascade with the previous published spectrum from \cite{tsiaras_pop} shows that both spectra are in excellent agreement (Fig.~\ref{fig:HST}). The band-averaged transit depth of \cite{tsiaras_pop} is 150~ppm (less than $2\,\sigma$) larger than the averaged depth we derive. This difference is consistent with the quoted uncertainties and can be explained by the large systematics and sparse time sampling of the \hst data, in combination with the different methods used to fit the baselines of the spectral light curves.

\subsection{Comparison with models}

\subsubsection{Retrieval outcomes}
\label{sec:Retrieval outcomes}

\begin{figure*}
    \centering
    \includegraphics[width=0.95\textwidth]{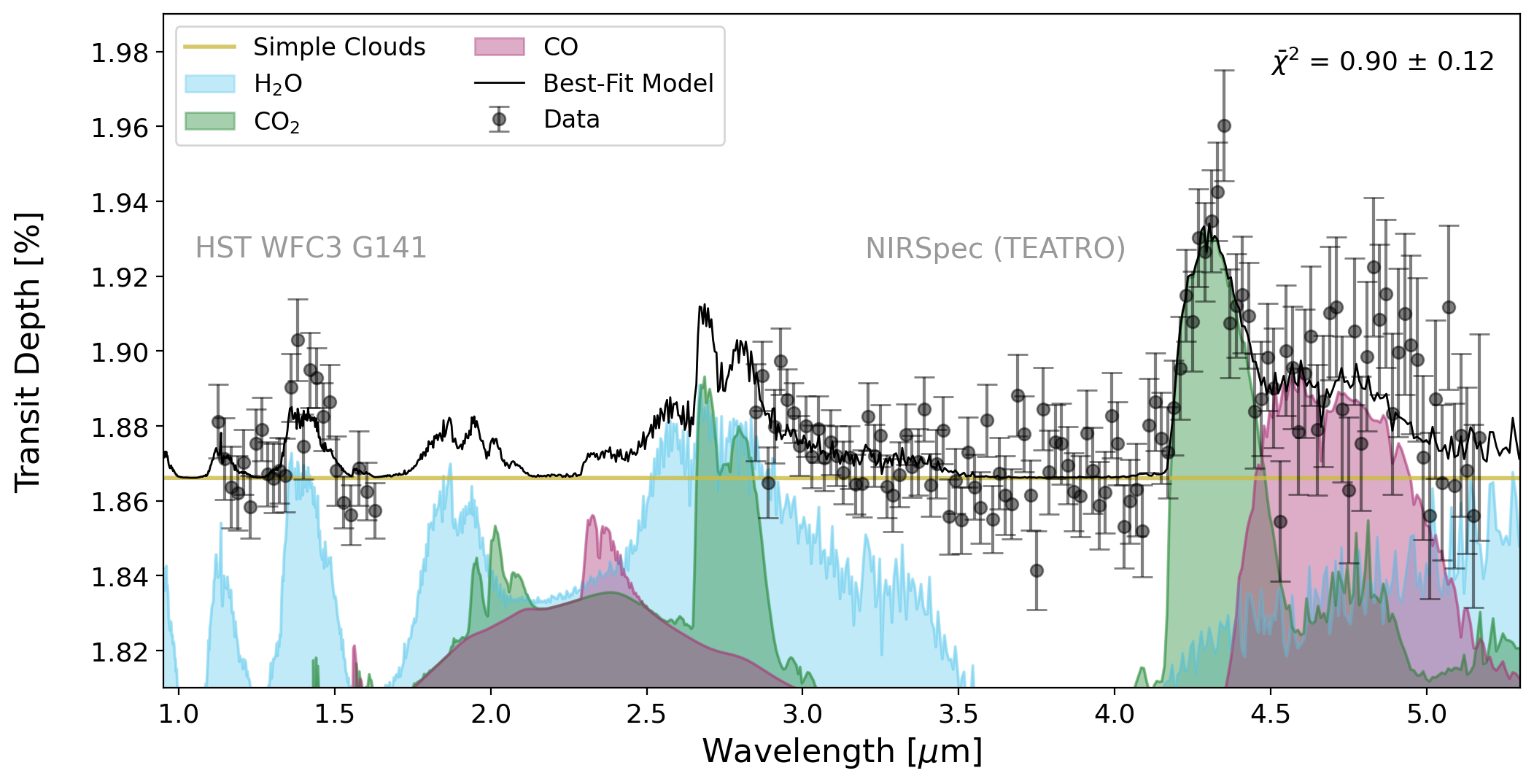}
    \caption{Best-fit model from the \arcis retrieval to the \teatro reduction of the \jwst NIRSpec G395M data and the \cascade reduction of the \hst WFC3 G141 data. Contributions from individual species are shown, as well as the reduced chi-square between the data and the best-fit model.}
    \label{fig:spectrum_w_model}
\end{figure*}

\begin{figure*}
    \centering
    \includegraphics[width=0.9\textwidth]{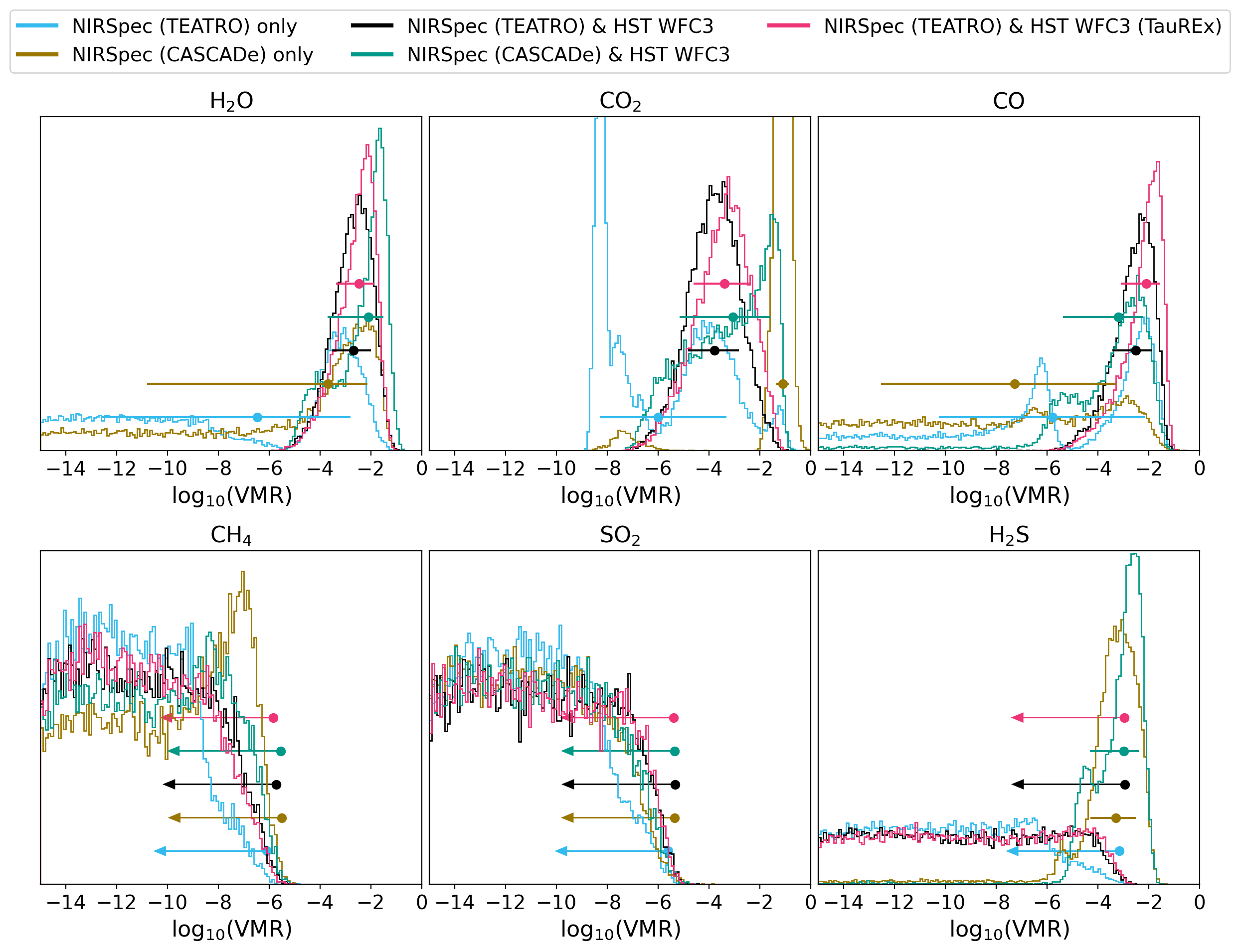}
    \caption{Probability distributions for different molecules for each retrieval conducted here. One-sigma constraints are also shown and arrows indicate $1\,\sigma$ upper bounds on the presence of a species.}
    \label{fig:abundances}
\end{figure*}

\begin{figure}
    \centering
    \includegraphics[width=\columnwidth]{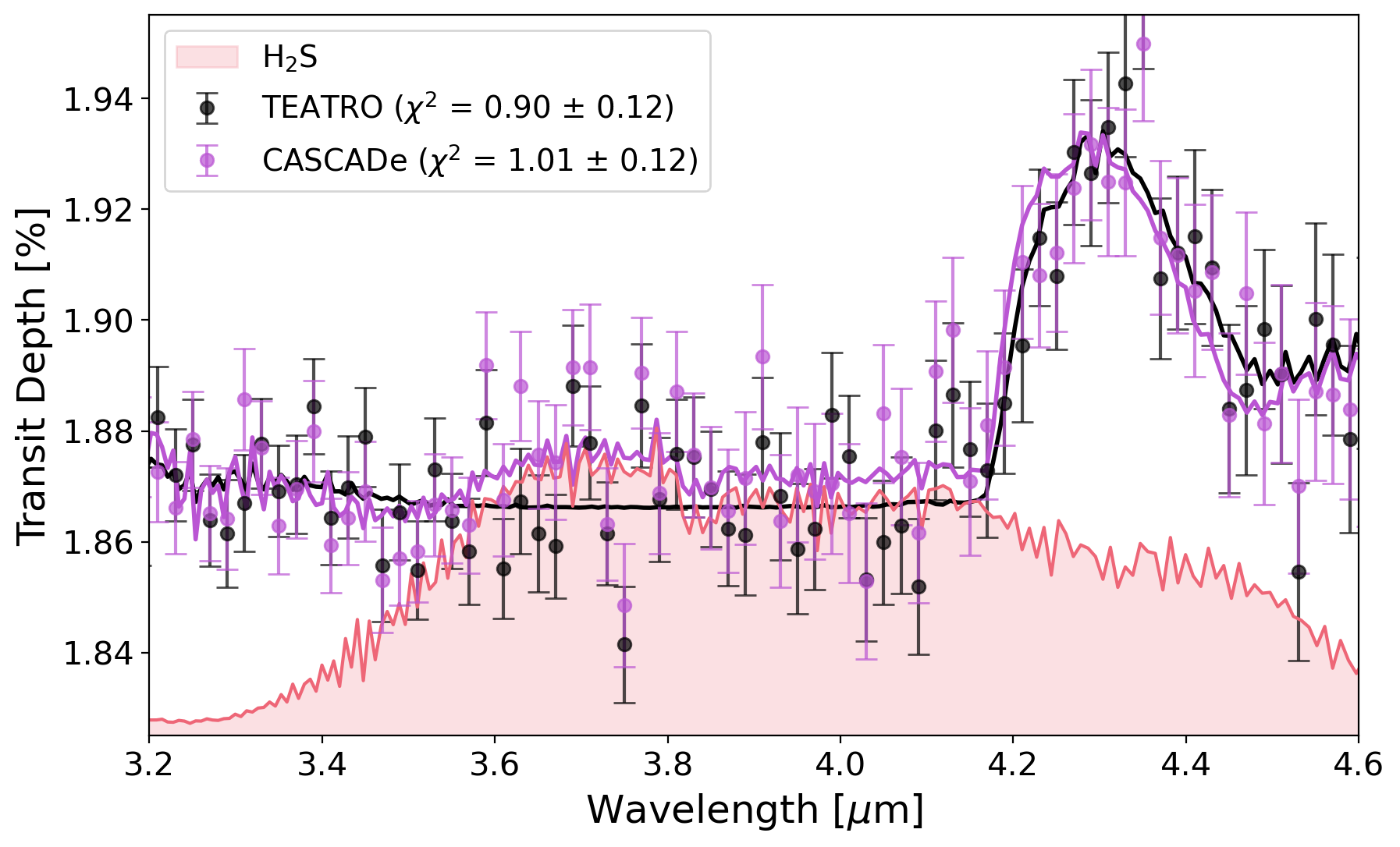}
    \caption{Best-fit models to each NIRSpec data reduction. The presence of H$_2$S is preferred for the \cascade reduction and the contribution of this molecule inferred from the retrieval analysis is shown. The reduced chi-squares between the data and the best-fit model are given.}
    \label{fig:spectrum-zoom-H2S}
\end{figure}

The best-fit model from the \arcis retrieval on the combined \hst WFC3 G141 and \jwst NIRSpec G395M (\teatro) spectra is presented in Fig.~\ref{fig:spectrum_w_model}, with the contributions of different molecules also given. Our atmospheric modelling finds strong evidence for three species: CO$_2$, CO, and H$_2$O. The probability distributions for these and the other species (CH$_4$, SO$_2$, and H$_2$S) are shown in Figs.~\ref{fig:abundances} and~\ref{fig:posteriors}; the latter also shows the planet radius at 10~bar, the atmospheric temperature, and the cloud pressure. The reduced chi-square between the data and best-fit model spectrum is $0.90 \pm 0.12$, which indicates that the model is an accurate representation of the data (Appendix~\ref{ap:chi-square}).
The retrieved abundances and detection significances are given in Table~\ref{tab:retrievals}. 
We find that CO$_2$, CO, and H$_2$O are detected at 12.2, 4.1, and 6.0~$\sigma$, respectively. From the \taurex cross-section retrieval, we find similar values (12.0, 3.7, and 5.9~$\sigma$, respectively), and these two retrievals are in excellent agreement overall. For the \arcis retrieval, but with the \cascade reduction of the NIRSpec data combined with the \hst WFC3 spectrum, we find similar abundances of these three molecules although their detection significances are slightly lower. The reduced chi-square between the data and best-fit model spectrum is $1.01 \pm 0.12$, so the model is also an accurate representation of the data. However, the fit to the \hst WFC3 and NIRSpec (\cascade) spectrum also prefers the presence of H$_2$S to $3.2\,\sigma$ (Fig.~\ref{fig:spectrum-zoom-H2S}). For the \teatro reduction, there is no evidence for H$_2$S and that part of the spectrum is fitted by clouds at lower pressures. 

For the retrievals on the NIRSpec data alone, we find that the constraints on CO$_2$, CO, and H$_2$O are far less robust, as evidenced by the varying constraints on the abundances and the large uncertainties (Figs.~\ref{fig:abundances} and \ref{fig:posteriors_nirspec_only}). 
The \ce{CO2} and CO posterior distributions from retrievals using NIRSpec alone are bimodal. 
This bimodality and the differences in the retrieved VMRs arise from a degeneracy between the abundances of CO and \ce{CO2} (i.e. the metallicity) and the presence of clouds. Indeed, the degeneracy disappears when clouds are removed from the models. It also disappears when the additional water band from \hst is included.

Using our base retrieval (\hst + \nirspec \teatro spectra and \arcis retrieval), we find an atmospheric temperature of $490^{+75}_{-60}$~K in the terminator region at the pressures probed by these observations ($P \in [4\times10^{-8} , 4\times10^{-3}]$~bar, Fig.~\ref{fig:contribution}). This temperature is similar to those derived from previous observations, for example $570^{+86}_{-78}$ and $456^{+70}_{-40}$~K found by \citet{Jiang2021} and \citet{Pinhas2019}, respectively. This points towards a relatively cold upper atmosphere for \mbox{HAT-P-12\,b}.
However, transmission spectrum retrievals using 1D models commonly find terminator temperatures that are significantly lower than the planet's equilibrium temperature and also lower than the terminator temperatures computed from GCMs \citep[\eg][]{Skaf2020, tsai_so2, Constantinou2023}. For \mbox{HAT-P-12\,b}, a GCM simulation by \citet{Kataria2016} that assumed a cloudless and solar metallicity atmosphere predicts east- and west-terminator-averaged temperatures around 800 and 600~K respectively at the pressures probed here. These differences may arise from a bias introduced by 1D models compared to GCMs \citep[\eg][]{Caldas2019, MacDonald2020, Pluriel2020}. The planets' albedos might also play a role. Disentangling the morning and evening terminator transmission spectra could help us understand the 1D retrieval outcomes, and has been done on \nirspec data of \mbox{WASP-39\,b} \citep{Espinoza2024}.
The retrieved planet radius at 10~bar is $0.89^{+0.01}_{-0.01}$~$\rm R_{Jup}$. The presence of clouds is preferred at $4.6\,\sigma$ with a cloud pressure range comprised between 2 and 11~mbar ($1\,\sigma$ bounds). With \cascade, the presence of clouds is less significant ($2.3\,\sigma$) and their pressure range is a bit higher (5--269~mbar).

\subsubsection{Forward model outcomes}
\label{subsec:forward models}

The initial optimisation procedure described in Sect.~\ref{subsec: forward models methods} leads to an atmosphere with high intrinsic temperature, quasi-solar carbon-to-oxygen ratio, high metallicity, and a moderate to strong vertical mixing coefficient (Table~\ref{tab:forward_model_opt}).

The derived $T_{\rm int}$ is 800~K and reaches the upper bound of the allowed range. It is driven by the lack of a methane feature in the spectrum and is a way for our models to decrease the methane abundance. It is very high compared to model predictions for hot Jupiters: $\sim$190~K would be expected for \mbox{HAT-P-12\,b} from Eq.~(3) of \citet{Thorngren2020, Thorngren2019}.
Such frameworks consider stellar irradiation as the main heat source, and other processes such as tidal heating or delayed cooling can add additional heat to the interior, so that the actual $T_\mathrm{int}$ can deviate from model predictions \citep[\eg][]{Jermyn2017, Sainsbury-Martinez2019, Fortney2021, Komacek2020, Sarkis2021, Thorngren2021}.
To compute the additional energy dissipated by the tides, we used Eq.~(23) of \citet{Leconte2010} with the star and planet parameters from Table~\ref{tab:system parameters}, an eccentricity of 0.026 \citep{Knutson2014}, the same ratio of Love number $k_{\rm2}$ to tidal quality factor $Q$ as Saturn \citep[$k_{\rm2}=0.39\pm0.024$, $k_{\rm2}/Q=(1.59\pm0.74)\times10^{-4}$,][]{Lainey2017}, and assumed that this energy is dissipated close to the surface of the planet. We find that $T_{\rm int}$ would reach $380-490$~K. Adding delayed cooling with the favourable assumption that it contributes to half the energy budget, $T_{\rm int}$ would reach $460-580$~K. Thus, it appears very difficult for these models to explain a $T_{\rm int}$ of 800~K. Other methane depletion mechanisms not included in our models may also be at play (Sect.~\ref{sec:Absence of methane}).

\begin{figure}
    \includegraphics[width=\columnwidth]{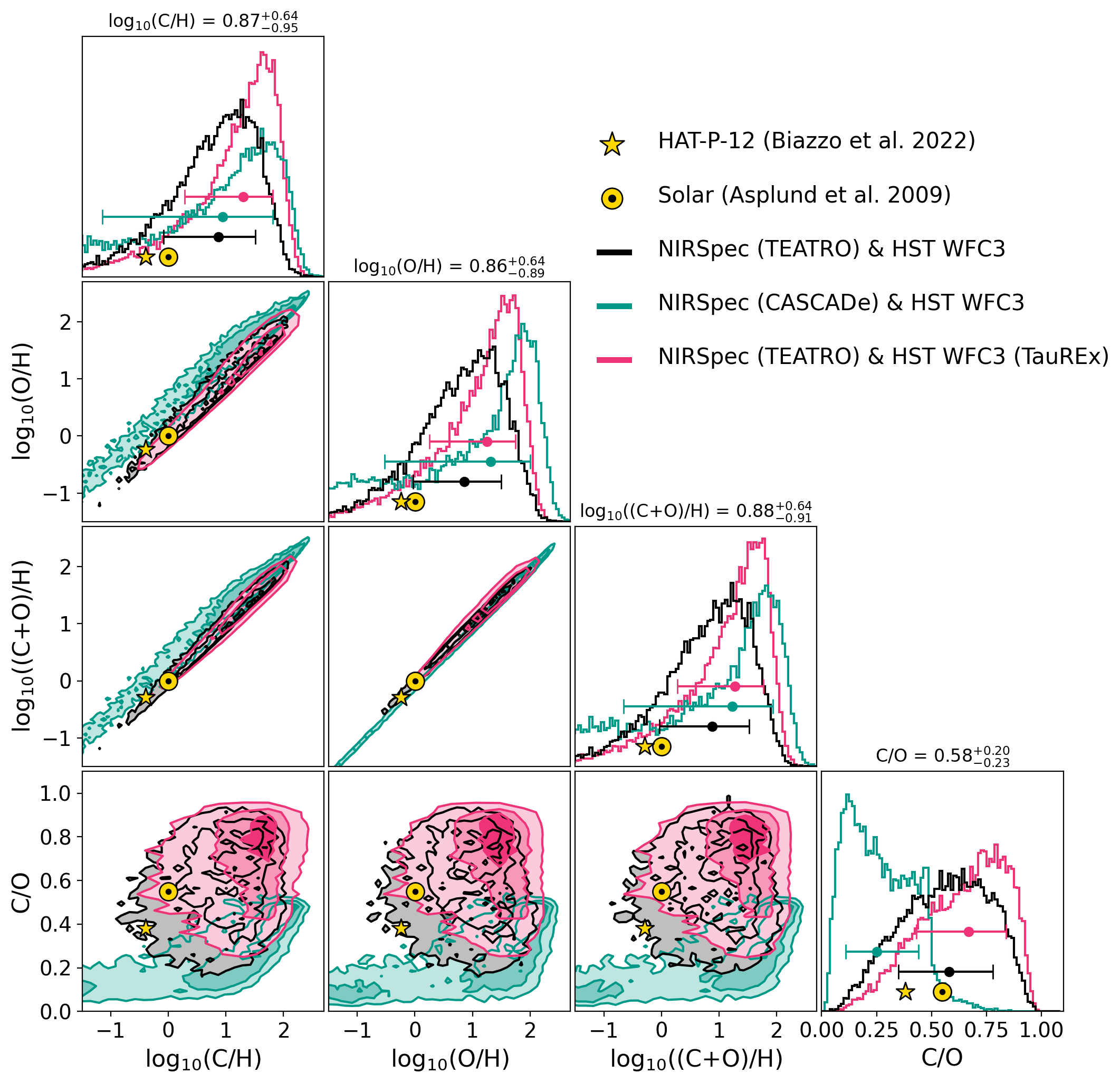}
    \caption{Elemental ratios derived in this work for HAT-P-12\,b in comparison to solar values and to those of the host star. The C/H, O/H, and (C+O)/H ratios are relative to solar.}
    \label{fig:ratios}
\end{figure}

\begin{figure*}
    \includegraphics[width=\textwidth]{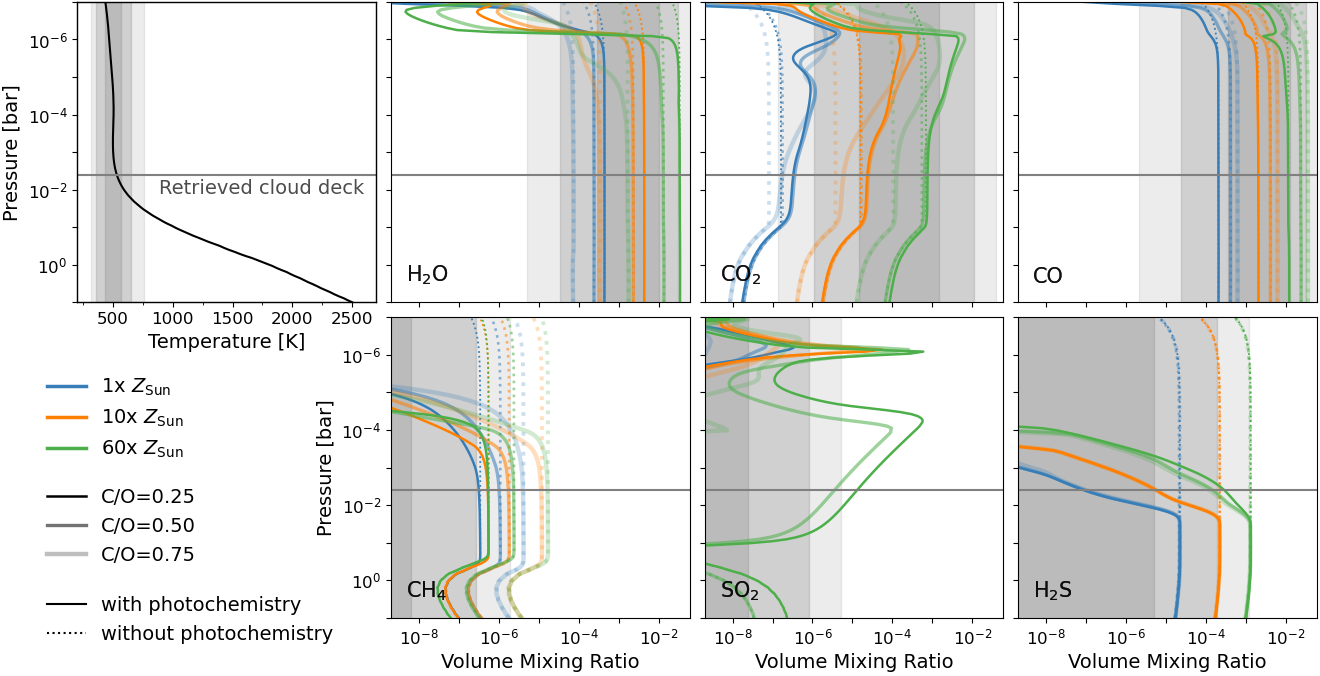}
        \caption{\vulcan chemistry results for HAT-P-12\,b. The top left panel shows the $P-T$ profile with high intrinsic temperature that results from the optimisation procedure described in Sect.~\ref{subsec: forward models methods}. The other panels show the VMRs of different molecules detected or expected in the NIRSpec spectrum, for different metallicities, carbon-to-oxygen ratios, and with or without photochemistry. From dark to light grey, the shaded areas indicate respectively the 1, 2, and $3\,\sigma$ uncertainty intervals of the retrieved atmospheric temperature and of the VMRs for detected species, and the 1, 2, and $3\,\sigma$ upper limits of the VMRs for non-detected species (inferred from the \hst + \nirspec \teatro spectra and \arcis retrieval).}
    \label{fig:vulcan}
\end{figure*}

\begin{figure}
    \includegraphics[width=\columnwidth]{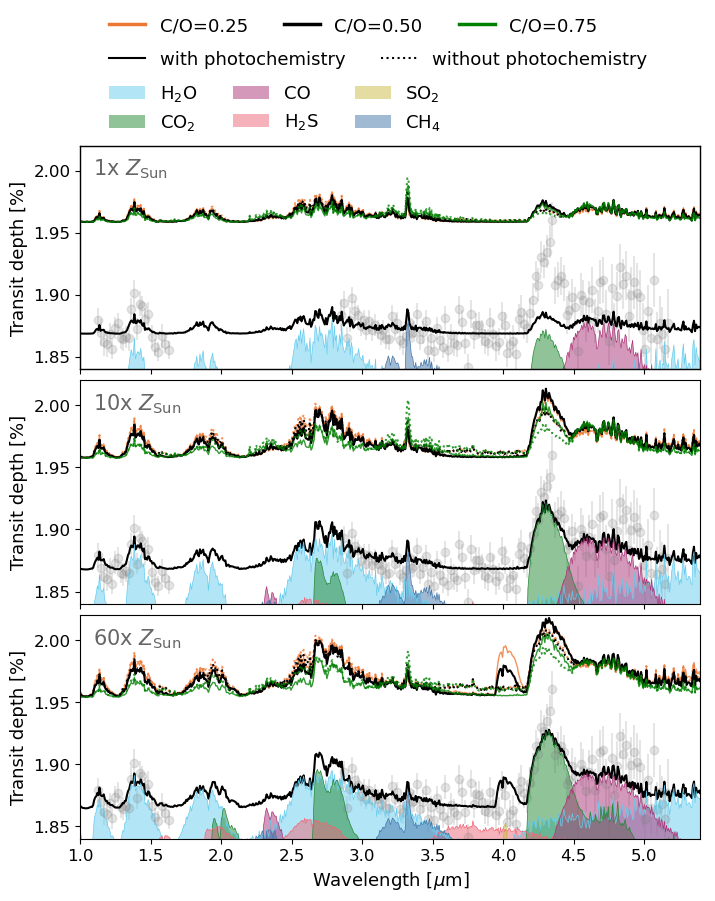}
    \caption{Synthetic spectra with grey clouds based on \vulcan chemistry models for \mbox{HAT-P-12\,b}. From top to bottom, the metallicity is 1, 10, and 60 times solar. Models with carbon-to-oxygen ratios of 0.25 (orange), 0.50 (black), and 0.75 (green) with photochemistry (solid lines) and without photochemistry (dotted lines) are shown. An offset has been applied to the reference radius-pressure points of the model spectra to match the observed NIRSpec transit depths, and an offset of $0.09\,\%$ has been added for visual purposes. The \jwst NIRSpec G395M data (\teatro reduction) and the \hst WFC3 G141 data are shown in light grey. The coloured regions correspond to the contribution of individual species to the model spectrum with C/O of 0.50 and photochemistry, which is also superimposed on the data.}
    \label{fig:vulcan_spectra}
\end{figure}

With the \vulcan models, we explored variations of the metallicity and C/O ratio as our initial optimisation did not take into account (photo)chemical kinetics, which can impact the resulting atmospheric composition.
We used metallicities of 1, 10 (retrieved value, Fig.~\ref{fig:ratios}), and 60 (optimisation, Table~\ref{tab:forward_model_opt}) times solar.
Given the differences in retrieval results between different datasets and reductions and the resulting spread in carbon-to-oxygen ratios (Figs.~\ref{fig:abundances} and \ref{fig:ratios}), we adopted C/O ratios of 0.25, 0.5, and 0.75. Finally, we computed the models with and without photochemistry, to highlight its effect on the atmospheric composition.
The \vulcan simulations show that a model with 10 times solar metallicity, solar or subsolar C/O, a $T_{\rm int}$ of at least 800~K, and photochemistry can explain most of the observed spectral features and the retrieved VMRs (Figs.~\ref{fig:vulcan} and \ref{fig:vulcan_spectra}). In that framework, part of the CO$_2$ is produced by photochemistry in the upper atmosphere. The production of \ce{SO2} is hampered by a relatively cold upper atmosphere; a higher temperature or higher metallicity would be required to produce an observable feature. Methane is largely destroyed via the hot intrinsic temperature and quenched by strong vertical mixing. Altogether, these findings provide physical insights to understand the observed spectrum, and are discussed in detail in Sect.~\ref{sec:Discussion}.
It should be noted that the parameters used in our \vulcan simulations, while obtained from an initial optimisation or set to values that are consistent with the retrievals, were not optimised to best fit the observed spectrum for each simulation. Thus, the comparison between the VMRs obtained from the retrievals and from \vulcan is meant to be informative but should be interpreted with caution.

\begin{table}
    \centering
    \caption{Best-fit parameters obtained from our forward model optimisation procedure.}
\renewcommand{\arraystretch}{1.2}
    \begin{tabular}{lccc} \hline \hline \\[-4.2mm]
Parameter   & Unit & Range allowed  & Value \\
    \hline \\[-4mm]
$R_{\rm p}$ & $\rm R_{Jup}$ &   0.5 - 1.4   &   0.87 \\
$\beta$     &            &   0 - 0.25    &   0.05 \\
$T_\mathrm{int}$ & K   &   50 - 800    &   800 \\
C/O              &       &   0.2 - 0.9   &   0.58 \\
Metallicity & $\rm Z_{\odot}$ &   1 - 100    &   63 \\
$K_{\rm zz}$ & $\rm cm^2\,s^{-1}$   &   $10^3$ - $10^{15}$  &   $1.6\times10^8$
  \\ \hline
    \end{tabular}
    \tablefoot{$\rm Z_{\odot}$ is the solar metallicity.}
    \label{tab:forward_model_opt}
\end{table}

\section{Discussion}
\label{sec:Discussion}

\subsection{Derived elemental ratios}
\label{sec:elemental_ratios}

We used our retrieved molecular abundances to derive three elemental ratios: C/O, C/H, and O/H. For these calculations, we used only the abundances of CO$_2$, CO, and H$_2$O as these are the only species which are confidently constrained. We performed this calculation for our \arcis retrieval on the combined \hst WFC3 G141 and \jwst NIRSpec G395M (\teatro) spectra. In Fig.~\ref{fig:ratios}, we compare our derived abundance ratios to those of the host star, \mbox{HAT-P-12}, which were reported in \citet{biazzo_stellar_abund}. We acquired these from the Hypatia catalogue \citep{hinkel_cat} and normalised them to the solar abundances from \citet{asplund_solar}. Hence, our results suggest that the abundances of carbon and oxygen in the atmosphere of the planet \mbox{HAT-P-12\,b} could be enriched with respect to the prevalence of the same species in the host star (i.e. the planet has a superstellar metallicity). 

From the \teatro reduction, the derived C/O ratio is also larger than the stellar value and is close to solar. However, using the \cascade reduction, the retrieved abundances are slightly different and the derived C/O ratio in this case is lower (the VMRs have $1\,\sigma$ uncertainties of about one order of magnitude so small differences in VMRs have a strong impact on the derived C/O). Given the strong correlations between these ratios, and thus the large uncertainties in their values, they should be interpreted with caution. Additionally, these data only probe a small part of the planet's atmosphere and therefore the derived values may not be representative of the bulk elemental ratios.

\begin{figure}
    \includegraphics[width=\columnwidth]{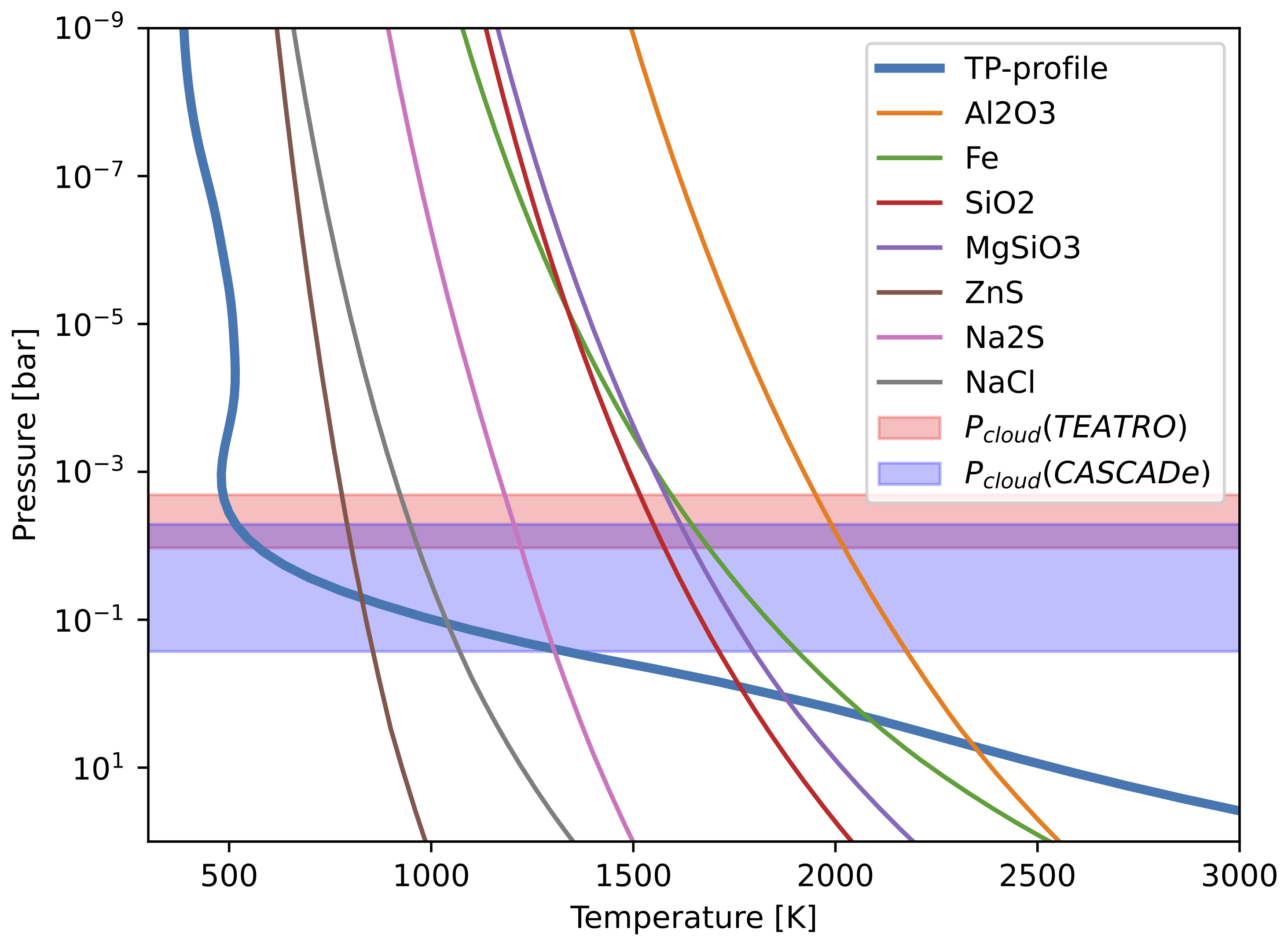}
    \caption{Condensation curves of several potential cloud-forming species (thin coloured lines) overplotted on the $P-T$ profile from the radiative-convective equilibrium model (thick blue line). The condensation curves assume a gas with $10\times$ solar metallicity. The shaded areas are the cloud pressure one-sigma ranges derived from the retrievals using the \teatro reduction (pink) and the \cascade reduction (blue).}
    \label{fig:condensates}
\end{figure}

Specifically, the elemental abundances in the upper part of the atmosphere are known to be impacted by selective rainout into cloud species. In Fig.~\ref{fig:condensates}, we show the condensation curves of several potential cloud species together with the $P-T$ profile derived from the radiative-convective equilibrium model (Sect.~\ref{subsec: forward models methods}). This plot suggests that at the cloud pressure levels inferred from the retrievals, these would be dominated by Na$_2$S, ZnS, and/or NaCl particles. Many more refractory species, such as silicates and several oxides, are expected to rain out below the observable atmosphere. Indeed, post-processing the model with cloud formation according to the concept from \citet{Ormel2019} extended for multiple species using the formalism of \citet{Huang2024}, without taking the radiative feedback of the clouds into account, we find that the upper clouds are composed of these three cloud species. It should be stressed that the radiative feedback of the clouds themselves might significantly alter the $P-T$ profile and by that the condensation pressure of the cloud species. Therefore, all interpretations based on the ratios derived here should be done cautiously.

\subsection{Abundances of \ce{CO2}, CO, and \ce{H2O}}
\label{sec:molecular_abundances}

The $P-T$ profile and resulting VMRs obtained with \vulcan are shown in Fig.~\ref{fig:vulcan}.
Models with solar metallicity produce too little \ce{CO2} by $2\,\sigma$ and $3\,\sigma$ with and without photochemistry, respectively, and too little CO and \ce{H2O} by $1\,\sigma$.
With a $10\times$ solar metallicity and no photochemistry, the CO and \ce{H2O} VMRs are well reproduced but the \ce{CO2} VMR is too low by $1\,\sigma$. 
Adding photochemistry increases the amount of \ce{CO2} by one order of magnitude in atmospheric layers probed by the observations and brings it in line with the retrieved value. This is predominantly due to its photochemical production via the reaction \ce{OH + CO -> CO2 + H}, where \ce{OH} radicals originate from water photolysis \ce{H2O ->[$+\,\nu$] OH + H} and thermolysis \ce{H2O + H -> OH + H2} (the relative importance of these two reactions depends on the pressure layer). Photochemistry affects only marginally the CO and \ce{H2O} VMRs for pressures above $10^{-7}$~bar.

Another way to increase the \ce{CO2} VMR to fit within $1\,\sigma$ is to turn off photochemistry and increase the metallicity to $60\times$ solar. However, with a C/O of 0.25 or 0.50, these simulations yield too much CO and/or \ce{H2O} by 1 to $2\,\sigma$.
Increasing the C/O ratio to 0.75 decreases the \ce{H2O} VMR enough to match the observation, and also decreases the \ce{CO2} VMR; however, the CO VMR is too high by $2\,\sigma$.
Thus, even with a supersolar C/O, such a metallicity as high as 60$\times$ solar is not preferred. This shows that the high metallicity found by our initial optimisation procedure (Table~\ref{tab:forward_model_opt}) is a direct consequence of not taking into account photochemical production of \ce{CO2}.
We note that the C/O retrieved from \teatro is consistent with 0.5 and 0.75 and that retrieved from \cascade is lower and consistent with 0.25; however, they have large uncertainties.
Overall, an atmosphere with a $10\times$ solar metallicity, which is close to that of Saturn \citep{Guillot2023}, and production of \ce{CO2} by photochemistry can best explain our observed abundances of \ce{CO2}, CO, and \ce{H2O}.

\ce{CO2} has been detected in the atmospheres of other gas giant exoplanets with \jwst (Sect.~\ref{sec:Introduction}) and its abundance has been used to constrain their atmospheric metallicities. However, to what extent the \ce{CO2} abundance is related to that of other compounds and to the overall atmospheric metallicity depends on the assumptions regarding atmospheric chemistry. This study shows that the photochemical production of \ce{CO2} can increase its abundance significantly and may need to be considered when deriving atmospheric metallicities.

\subsection{Sulphur-bearing species}

Several studies have found evidence for SO$_2$, a product of photochemistry, in gas giant exoplanet atmospheres (Sect. \ref{sec:Introduction}). 
In this work, we do not see evidence for this molecule. Instead, the retrievals can fit the data without \ce{SO2} spectral features, and tentatively with \ce{H2S} spectral features. Photolysis can form \ce{SO2} by destroying \ce{H2S}, so in the absence of \ce{SO2}, \ce{H2S} should remain in the atmosphere and should be the dominant sulphur reservoir \citep[\eg][]{polman_h2s_so2,tsai_so2}. However, \ce{H2S} is difficult to observe in the NIRSpec spectral range under the conditions of \mbox{HAT-P-12\,b}, as shown by the synthetic spectra in Fig.~\ref{fig:vulcan_spectra}.

The \vulcan simulations show that the production of \ce{SO2} depends on the temperature in the upper layers, the metallicity, the C/O ratio, and the surface gravity when comparing different planets.
The dominant factor to explain the non-detection of \ce{SO2} for \mbox{HAT-P-12\,b} in the NIRSpec range (around 4~\mic, Fig.~\ref{fig:vulcan_spectra}) seems to be the upper atmosphere temperature. Based on the chemical pathways of sulphur species described in \citet{tsai_so2}, we investigated different reactions that lead to \ce{SO2} formation. 
Consistent with the findings of \citet{tsai_so2}, we find that the reaction that converts \ce{H2O} into OH radicals via thermal dissociation (\ce{H2O + H <-> OH + H2}) is slower by one to four orders of magnitude at our retrieved temperature of 500~K compared to simulations with warmer upper layers of 600 to 800~K. This puts a significant constraint on the \ce{OH} availability for subsequent \ce{SO2} formation (Fig.~\ref{fig:SO2temperature}).
On the other hand, we find that \ce{H2O} photodissociation is efficient over a broader pressure range in the 500~K atmosphere model compared to those at $700 - 800$~K, up to pressures of $10^{-3}$ and $10^{-5}$~bar respectively (for models with a 10$\times$ solar metallicity, a C/O of 0.5, with photochemistry), because the UV radiation penetrates to higher pressures and there is more \ce{H2O} available.
In the 500~K atmosphere model, the main OH source is \ce{H2O} photolysis at pressures below $10^{-5}$~bar, \ce{H2O} thermolysis slightly dominates over photolysis at pressures between $10^{-5}$ and $10^{-3}$~bar (although it varies with pressure), and thermolysis dominates at pressures above $10^{-3}$~bar.
Some of the OH that is freed by \ce{H2O} photolysis reacts with \ce{H2} to form \ce{H2O} again.
We find that \ce{SO2} is produced from OH and \ce{H2S} but not in a quantity that is enough to be detected. A warmer temperature would increase the amount of available OH to form \ce{SO2}.
Considering the potential bias in 1D models (Sect.~\ref{sec:Retrieval outcomes}), upper atmospheric temperatures higher by 200 or 300~K (i.e. 700 or 800~K) would increase the \ce{SO2} VMR and bring it close to its 1.5 or $2\,\sigma$ detection limit, respectively (Figs.~\ref{fig:SO2temperature} and~\ref{fig:vulcan}), thus these higher temperatures are also marginally consistent with the non-detection of \ce{SO2}. According to these models, a temperature above $\sim$800~K would be required to be able to detect \ce{SO2} at 4~\mic.

Regarding metallicity, simulations including photochemistry show that a solar or $10\times$ solar metallicity does not produce a significant \ce{SO2} feature.
A higher metallicity (\eg $60\times$ solar) and a solar or subsolar C/O ratio are required to have enough available oxygen to potentially produce a detectable \ce{SO2} feature, with a VMR around its retrieved $3\,\sigma$ upper limit (Fig.~\ref{fig:vulcan} and~\ref{fig:vulcan_spectra}).
The fact that \ce{SO2} is not detected in the observed spectrum might be another indication that a $60\times$ solar metallicity is too high, although a high C/O ratio (\eg 0.75) is also a possible explanation as it suppresses the \ce{SO2} feature even with a $60\times$ solar metallicity.
At such a high C/O, most oxygen is locked up in \ce{CO}, which reduces its availability for \ce{SO2}.
However, this combination of high metallicity and high C/O conflicts with retrieved abundances for CO and \ce{CH4}.
These results differ from \mbox{WASP-107\,b} where \ce{SO2} was prominent at $10\times$ solar metallicity \citep{dyrek_w107}, but the retrieved temperature was in the range $720 - 850$~K and its surface gravity is 1.8 times lower \citep[which increases the amplitudes of spectral features observed in transmission, but also decreases the pressures at which the UV radiation can penetrate, see][]{Baeyens2022}.

In our synthetic spectra in Fig.~\ref{fig:vulcan_spectra}, \ce{H2S} does not contribute strongly to the NIRSpec spectrum (even without photochemistry), and the different inference on \ce{H2S} from both data reductions precludes from providing a strong constraint on this species. In these models, the \ce{H2S} feature around 3.8~\mic starts being noticeable at high metallicities (at least $10\times$ solar). 
Further data with MIRI LRS, which probes spectral regions that contain stronger H$_2$S and SO$_2$ features \citep{polman_h2s_so2,dyrek_w107} should provide stricter constraints on the presence of these species (Bouwman et al. in prep.).

\subsection{Absence of methane}
\label{sec:Absence of methane}

Across all our atmospheric retrievals, no evidence is found for CH$_4$. Instead, each places an upper bound on the abundance of this molecule. For our standard retrieval, the $3\,\sigma$ upper bound of the abundance of methane is log$_{10}$(VMR)~=~$-5.73$ and the other joint \hst and \jwst retrievals give similar values (Fig.~\ref{fig:abundances}, Table~\ref{tab:retrievals}). Given the planet's equilibrium temperature and its apparent supersolar metallicity (Sect.~\ref{sec:elemental_ratios}), one would expect methane to be prevalent throughout the atmosphere. If the planet's atmosphere were in chemical equilibrium, the methane abundance could be diminished by a low metallicity, a high temperature in the upper layers, or a low C/O ratio. A low metallicity or a high temperature would contrast with the results of our retrievals (Figs.~\ref{fig:ratios} and~\ref{fig:posteriors}). The retrieved upper atmospheric temperature ($\sim$500~K) would have to be severely underestimated (Sect.~\ref{sec:Retrieval outcomes}) if the actual temperature were high enough ($\gtrsim$~1000~K) to explain the low methane abundance. Therefore, the non-detection of methane may be suggestive of disequilibrium chemistry in the atmospheric layers probed by these data. The retrievals find a wide range of possible C/O ratios, thus a potentially low value (\eg 0.25 or even lower as found by the retrieval from \cascade) could also play a role.

In the \vulcan simulations, we adopted a moderate to high eddy diffusion coefficient $K_{\rm zz}$ and a $P-T$ structure with high intrinsic temperature to lower the available methane in the quenched atmosphere \citep{Fortney2020, dyrek_w107}. We find that \ce{CH4} is photochemically destroyed in the high atmosphere by reacting with hydrogen (\ce{H + CH4 -> CH3 + H2}). Nevertheless, a methane feature remains in the simulated spectra (Fig.~\ref{fig:vulcan_spectra}) and its abundance is on the high end of the retrieval results: for C/O ratios of 0.25 and 0.5, its VMR is around the $2\,\sigma$ and $3\,\sigma$ upper limit obtained from the retrievals, respectively, and for a C/O of 0.75 it is higher than the $3\,\sigma$ upper limit (Fig.~\ref{fig:vulcan}). A lower $T_{\rm int}$ (Sect.~\ref{subsec:forward models}) would increase the methane abundance. On the other hand, unfavourable noise may hinder its detection. Overall, methane seems to be lacking.
As a comparison, methane was detected in other warm gas giants: \mbox{WASP-107\,b} with a VMR of $10^{-6}$ that is close to the $3\,\sigma$ upper limit we find for \mbox{HAT-P-12\,b}, and \mbox{WASP-80\,b} with a larger VMR of $10^{-4}$ \citep{Sing2024, Welbanks2024, bell_w80}.

Mechanisms not included in our \vulcan simulations seem necessary to explain the apparent depletion of methane.
The $K_{\rm zz}$ parametrisation assumes that atmospheric mixing mimics diffusion. In our \vulcan models, \ce{CH4} is quenched at pressures around one bar. High intrinsic temperatures can push the radiative-convective boundary to such low pressures \citep[\eg][]{Thorngren2019} and convective mixing could quench \ce{CH4} even deeper in the atmosphere. Including convective mixing in future models would help us determine the causes of the \ce{CH4} depletion; a first step could be to increase $K_{\rm zz}$ at the radiative-convective boundary.
Horizontal mixing from the equatorial jet stream could also push \ce{CH4} towards the dayside, where higher temperatures would facilitate its elimination. This mechanism has been invoked to explain the lack of methane on the nightside of \mbox{WASP-43\,b} \citep{Bell2024}. Our models are 1D and do not include horizontal mixing. The spectrum shown here probes the terminator regions, both morning and evening, and \mbox{HAT-P-12\,b} has a lower $T_{\rm eq}$ than \mbox{WASP-43\,b} ($\sim$955~K and $\sim$1400~K, respectively). Simulations by \citet{Baeyens2021, Baeyens2022b} based on a large grid of GCMs and pseudo-2D chemical kinetics show that such horizontal mixing has little impact on the overall \ce{CH4} abundance for planets below $T_{\rm eff}\sim1400$~K, whereas a reduction of \ce{CH4} occurs above that $T_{\rm eff}$. Thus, we expect horizontal mixing to be a secondary effect for \mbox{HAT-P-12\,b}. 
Our results can be compared to those obtained for \mbox{WASP-107\,b}, where the low methane abundance points to a high interior temperature in combination with strong vertical mixing \citep{dyrek_w107, Sing2024, Welbanks2024}. For \mbox{WASP-107\,b}, the 1D approach is also justified because the day$-$night temperature gradient and resulting jet stream for $T_{\rm eff}\sim750$~K planets are expected to be small. These hot interior temperatures could also contribute to a planet's inflation \citep[\eg][]{Sainsbury-Martinez2019, Komacek2020, Sarkis2021}.

Our retrievals indicate the presence of clouds at low pressures. In their study on the role of clouds on the depletion of methane in the atmospheres of irradiated exoplanets, \citet{Molaverdikhani2020} found that methane is depleted rather than obscured by clouds. This depletion is caused by excess heating by clouds that releases oxygen from condensates and accelerates the formation of water instead of methane, the carbon being captured to form \ce{CO2} and CO. In their cloudy and cold atmosphere models ($T$~=~800~K), methane vanishes at relatively low C/O ratios ($\lesssim$~0.6). For a solar metallicity, this requires a low sedimentation factor ($f_{\rm sed}\lesssim$~1), which is consistent with the ${\rm log}(f_{\rm sed}) = -0.98^{+0.22}_{-0.30}$ reported by \citet{Yan2020} for \mbox{HAT-P-12\,b}. This heating is also more efficient when the metallicity is high. Given the tentative close-to-solar C/O ratio and supersolar metallicity inferred for the atmosphere of \mbox{HAT-P-12\,b}, this process may play a role in the non-detection of methane, and potentially in that of \mbox{WASP-107\,b} where silicate clouds have been detected \citep{dyrek_w107}.

The NIRISS SOSS wavelength range (0.6--2.8~\mic) contains several potential \ce{CH4} bands, so the \mbox{HAT-P-12\,b} NIRISS spectrum should provide further constraints on the abundance of this species (Heinke et al. in prep.).

\subsection{Formation of HAT-P-12\,b}

Comparing the metallicity and C/O ratio of planets and exoplanets to those of their host star provides clues to constrain their formation in the protoplanetary disk \citep[\eg][]{Oberg2011, Fortney2013, Madhusudhan2017, Molliere2022}.
The tentative superstellar metallicity of \mbox{HAT-P-12\,b} may indicate that the accretion was dominated by solids, which should yield a substellar C/O ratio \citep[\eg][]{Oberg2011, Madhusudhan2014, Madhusudhan2017, Mordasini2016}.
Alternatively, pebble drift and evaporation can lead to superstellar enrichment of the gas, and to C/O ratios that can be superstellar \citep{Booth2017, Schneider2021a, Schneider2021b, Molliere2022}. As noted in \citet{Molliere2022}, superstellar enrichment and superstellar C/O ratio are difficult to explain without considering pebble evaporation. A superstellar metallicity and superstellar C/O ratio for \mbox{HAT-P-12\,b} (as retrieved from \teatro but not from \cascade) would indicate that pebble evaporation may have played a role during its formation.
The composition and transport of cold planetesimals in the protosolar nebula have also been extensively studied to explain the supersolar metallicity of Jupiter and Saturn, and formation scenarios that result in supersolar C/O ratios have been developed \citep[Sect.~4 of][and references therein]{Cavalie2024}.
Tighter measurements of the metallicity and C/O ratio of \mbox{HAT-P-12\,b} are necessary to investigate which mechanisms may have been dominant in its formation.

\subsection{Information content of NIRSpec G395M}

Our retrievals based solely on the \jwst NIRSpec G395M spectrum suggest that data from this instrument and grating alone are not sufficient to robustly determine the chemical composition of \mbox{HAT-P-12\,b}. While the CO$_2$ feature is clearly visible by eye, the retrievals on the \teatro and \cascade reductions do not properly constrain the abundances and the results greatly differ between them (Figs.~\ref{fig:abundances} and \ref{fig:posteriors_nirspec_only}).
Previous studies have shown the presence of H$_2$O in the atmosphere of \mbox{HAT-P-12\,b} \citep[\eg][]{wong_h12,tsiaras_pop,edwards_hst_pop}. Both CO and H$_2$O are strongly detected and their abundances are consistent once the \hst spectrum is included in the retrieval. On the other hand, the \ce{H2S} discrepancy remains. These results suggest that having spectral data at shorter wavelengths, in addition to NIRSpec G395M data, is crucial to constrain the chemical composition in the presence of clouds or other opacities (\eg H$_2$S) which can attenuate the retrieved H$_2$O feature at 3.5--4.0~\mic \citep[see also][]{Constantinou2023}. Nevertheless, in other cases, \hst WFC3 data can have a less drastic impact on the retrieved water abundance \citep[\eg][]{w77_hst_jwst}.

\section{Conclusion}
\label{sec:Conclusion}

This study presents the near-infrared transmission spectrum of \mbox{HAT-P-12\,b} obtained with \jwst NIRSpec G395M. By analysing it in combination with the \hst WFC3 G141 spectrum, we found strong evidence for three species: CO$_2$, CO, and H$_2$O. We placed constraints on their abundances and interpreted them in the light of chemical model simulations. 
Our analysis suggests that \mbox{HAT-P-12\,b} has elemental abundances of C and O greater than the stellar values, while the C/O ratio is not well constrained.
When including photochemistry in the chemical models, the amount of \ce{CO2} increases by one order of magnitude in the pressure range probed by these observations, and a 10$\times$ solar metallicity is enough to explain its mixing ratio; a very high metallicity (\eg 60$\times$ solar) is no longer necessary. This shows that metallicities derived from near-infrared spectra showing \ce{CO2} may need to account for its photochemical production pathways. This may impact further studies exploring mass-metallicity trends, linking exoplanet atmospheres to their interior properties, and to their formation history \citep{Mordasini2016, Molliere2022, Bean2023, edwards_hst_pop}.

The two reductions of the NIRSpec data led to differing results for H$_2$S. Atmospheric modelling on one reduction tentatively suggests the presence of this species, while the other does not. In both cases the best-fit model is an accurate representation of the data, with a reduced chi-square close to unity. The cause of this discrepancy has not yet been identified and is left for further investigation.
Methane was not detected and all retrievals rule out high abundances of this molecule, which could be due to a hot interior temperature and strong vertical mixing, although the required internal heat would be difficult to explain. A low C/O ratio (e.g. $<$~0.25) could also decrease the abundance of methane. Photolysis as modelled by \vulcan seems insufficient. Overall, additional mechanisms could be necessary to fully explain the methane depletion.
SO$_2$ was not detected, and we find that a low upper atmosphere temperature limits the production of this species (although the low temperature found by the 1D retrievals should be taken with caution). Other factors could reduce the amount of \ce{SO2}, such as a low metallicity, no photochemistry (both of which are not favoured by our analysis), or a high C/O ratio (which would make the methane depletion even harder to explain). The presence of clouds is inferred at pressures from a few to a few hundred millibars depending on the data reduction employed; these clouds could be dominated by \ce{Na2S}, ZnS, or NaCl particles. Clouds may also play a role in the non-detection of methane.

This study will be continued after including the upcoming \jwst MIRI and NIRISS transmission spectra of \mbox{HAT-P-12\,b}: MIRI LRS should provide tighter constraints on the presence of \ce{SO2} and \ce{H2S}, and NIRISS SOSS on the presence of methane, water, and clouds (Bouwman et al. in prep., Heinke et al. in prep.). Extracting the morning and evening terminator transmission spectra from the \nirspec data would show how the atmospheric properties vary between the two limbs, and may shed light on the 1D retrieval results obtained here. Altogether, these spectra should provide tighter constraints on the metallicity, elemental abundances, C/O ratio, and chemical processes at play in the atmosphere of \mbox{HAT-P-12\,b}, and would help us understand the formation mechanisms of the planet.

\begin{acknowledgements}

TK acknowledges funding from the KU Leuven Interdisciplinary Grant (IDN/19/028).
P.-O.L., A.B., C.C., A.C., R.G., and D.R. acknowledge funding support from CNES.
JPP acknowledges financial support from the UK Science and Technology Facilities Council, and the UK Space Agency.
This project has received funding from the European Union's Horizon 2020 research and innovation programme under the Marie Sklodowska-Curie grant agreement no. 860470 (MC-ITN CHAMELEON).
MPIA acknowledges support from the Federal Ministry of Economy (BMWi) through the German Space Agency (DLR) and the Max Planck Society.
BV, OA, IA, and PR thank the European Space Agency (ESA) and the Belgian Federal Science Policy Office (BELSPO) for their support in the framework of the PRODEX Programme.
OA is a Senior Research Associate of the Fonds de la Recherche Scientifique – FNRS.
DB is supported by Spanish MCIN/AEI/10.13039/501100011033 grant PID2019-107061GB-C61 and No. MDM-2017-0737.
LD acknowledges funding from the KU Leuven Interdisciplinary Grant (IDN/19/028), the European Union H2020-MSCA-ITN-2019 under Grant no. 860470 (CHAMELEON) and the FWO research grant G086217N.
GO acknowledges support from the Swedish National Space Board and the Knut and Alice Wallenberg Foundation.
PP thanks the Swiss National Science Foundation (SNSF) for financial support under grant number 200020\_200399.
LC acknowledges support by grant PIB2021-127718NB-100 from the Spanish Ministry of Science and Innovation/State Agency of Research MCIN/AEI/10.13039/501100011033.
Support from SNSA is acknowledged.
TPR acknowledges support from the ERC through grant no.\ 743029 EASY.
EvD acknowledges support from A-ERC grant 101019751 MOLDISK.
This work is based on observations made with the NASA/ESA/CSA \textit{James Webb} Space Telescope. The data were obtained from the Mikulski Archive for Space Telescopes at the Space Telescope Science Institute, which is operated by the Association of Universities for Research in Astronomy, Inc., under NASA contract NAS 5-03127 for JWST. These observations are associated with program \#1281.
This research is based on observations made with the NASA/ESA \textit{Hubble} Space Telescope obtained from the Space Telescope Science Institute, which is operated by the Association of Universities for Research in Astronomy, Inc., under NASA contract NAS 5–26555. These observations are associated with program 14260.
\end{acknowledgements}

\bibliography{main}

\newpage

\begin{appendix}
 
\onecolumn

\section{System parameters}

\begin{table*}[h]
	\begin{center}
	\caption{Parameters of the HAT-P-12 system obtained from the literature and from the white light curve fits with both reduction methods.}
    \renewcommand{\arraystretch}{1.1}
    \begin{tabular}{p{0.16\textwidth}>{\centering}p{0.09\textwidth}>{\centering}p{0.25\textwidth}>{\centering}p{0.25\textwidth}>{\centering\arraybackslash}p{0.03\textwidth}}
	\hline
	\hline\\[-3mm]
	Parameter &  Unit  &  \multicolumn{2}{c}{Value}  &  Ref. \\[1mm]
	\hline\\[-3mm]
	\textbf{Star}  &   &   &   &  \\[1mm]
        Gaia DR3 source ID &  & \multicolumn{2}{c}{1499514786891168640} & 1, 2 \\
        RA   &   J2016.0  &  \multicolumn{2}{c}{13:57:33.269}      & 1, 2   \\
        Dec  &   J2016.0  &  \multicolumn{2}{c}{$+$43:29:35.894}   & 1, 2 \\
        $d$  & pc  &  \multicolumn{2}{c}{141.1859}    &  1, 2  \\
        K mag &    &     \multicolumn{2}{c}{$10.108 \pm 0.016$}    &     3  \\
        $T_{\rm eff}$  & K  &  \multicolumn{2}{c}{$4665 \pm 45$}   & 4 \\
        $\log{g_{\star}}$ & [g cm$^{-2}$] &   \multicolumn{2}{c}{$4.614 \pm 0.012$} & 4 \\
        $[$Fe/H$]$  & dex  &  \multicolumn{2}{c}{$-0.20 \pm 0.09$}  &  4 \\
        $M_{\star}$ & $\rm M_{\odot}$ &  \multicolumn{2}{c}{$0.691 \pm 0.032$}  & 4\\
        $R_{\star}$ & $\rm R_{\odot}$  &  \multicolumn{2}{c}{$0.679 \pm 0.012$} & 4 \\
        $u_1$ & & \multicolumn{2}{c}{$0.123 \pm 0.023$} & 5 \\
        $u_2$ & & \multicolumn{2}{c}{$ 0.179 \pm 0.050$} & 5 \\[1mm]     
    \hline \\[-3mm]
    \textbf{Planet}  &  & \teatro &  \cascade &   \\[1mm]
        $P$ & day & 3.21305751 & 3.21305992 & 6, 4 \\
        $T_0$ & BJD (TDB) & $2459987.425609 \pm 0.000023$  & -- &  \\
        $a/R_{\star}$ &  & 11.93 & 11.72 & 4, 7 \\
        $R_{\rm p}/R_{\star}$ &  & $0.13657 \pm 0.00022$ & $0.13730 \pm 0.00006$ &  \\
        $a$   & au &  0.03767  & 0.037 & 4, 7 \\
        $R_{\rm p}$ & $\rm R_{Jup}$ &   $0.9024 \pm 0.0014$ & $0.9072 \pm 0.0004$ &  \\
        $b$   &     &  $0.082 \pm 0.022$ &  0.225  &  \\
        $i$   & deg & $89.62 \pm 0.10$  &  88.9 &   \\
        $M_{\rm p}$ & $\rm M_{Jup}$ &  \multicolumn{2}{c}{$0.201 \pm 0.011$}  & 4 \\
        $T_{\rm eq}$ & K &   \multicolumn{2}{c}{$955 \pm 11$}  & 4 \\[0.5mm]
    \hline
    \hline
	\end{tabular}
    \tablefoot{
    1. \citet{Gaia2016}
    2. \citet{Gaia2023}
    3. \citet{Cutri2003}
    4. \citet{Mancini2018}
    5. This work (\teatro)
    6. \citet{Kokori2022}    
    7. This work (\cascade).
    $d$ is the distance. The limb-darkening coefficients $u_1$, $u_2$ are in the NIRSpec G395M bandpass. $b$ is the impact parameter. $a/R_{\star}$ was fixed to the value from \citet{Mancini2018} for \teatro and was a free parameter for \cascade, which causes some differences in the derived parameters. The uncertainties reported for \teatro and \cascade are only those from our fits, they do not take into account the uncertainties of the fixed parameters. The parameters from \cascade were obtained from an optimisation process that does not provide uncertainties (except for the transit depth), but are consistent with \citet{Mancini2018}. The \cascade fit used the mid-transit time from \teatro.}
    \end{center}
	\label{tab:system parameters}
\end{table*}

\FloatBarrier

\newpage

\section{Transmission spectrum}

\begin{table}[h]
	\begin{center}
	\caption{Transmission spectrum of HAT-P-12\,b in the 2.84--5.18~\mic range extracted with both reduction methods.}
    \vspace{-3.1mm}
\resizebox{0.9\textwidth}{!}{
\renewcommand{\arraystretch}{1.1}
	\begin{tabular}{ccccc}
	\hline
	\hline \\[-3mm]
	   & \multicolumn{2}{c}{\teatro} & \multicolumn{2}{c}{\cascade} \\[0.5mm]
	  Wavelength   &  Depth  &  Uncertainty  &  Depth  &  Uncertainty \\
    $[\mu\rm m]$   & [\%] & [\%] & [\%] & [\%] \\[1mm]
	\hline \\[-3mm]
2.85  &  1.8837  &  0.013  &  1.9112  &  0.0112  \\ 
2.87  &  1.8934  &  0.0093  &  1.9011  &  0.0107  \\ 
2.89  &  1.8648  &  0.0095  &  1.8783  &  0.0105  \\ 
2.91  &  1.8799  &  0.0099  &  1.8763  &  0.0107  \\ 
2.93  &  1.8973  &  0.0088  &  1.9105  &  0.0094  \\ 
2.95  &  1.8871  &  0.0081  &  1.8822  &  0.0095  \\ 
2.97  &  1.8836  &  0.008  &  1.8861  &  0.0091  \\ 
2.99  &  1.8747  &  0.008  &  1.873  &  0.0102  \\ 
3.01  &  1.88  &  0.0081  &  1.8778  &  0.0097  \\ 
3.03  &  1.8719  &  0.0083  &  1.8736  &  0.009  \\ 
3.05  &  1.8792  &  0.0088  &  1.8752  &  0.0092  \\ 
3.07  &  1.8714  &  0.0087  &  1.8724  &  0.0085  \\ 
3.09  &  1.8756  &  0.0087  &  1.885  &  0.0087  \\ 
3.11  &  1.872  &  0.0079  &  1.868  &  0.0081  \\ 
3.13  &  1.8676  &  0.0083  &  1.8758  &  0.0081  \\ 
3.15  &  1.8716  &  0.0084  &  1.8638  &  0.0085  \\ 
3.17  &  1.8645  &  0.008  &  1.876  &  0.0083  \\ 
3.19  &  1.8646  &  0.0089  &  1.8771  &  0.0091  \\ 
3.21  &  1.8825  &  0.0091  &  1.8726  &  0.009  \\ 
3.23  &  1.8721  &  0.0083  &  1.8662  &  0.0083  \\ 
3.25  &  1.8775  &  0.0083  &  1.8785  &  0.0086  \\ 
3.27  &  1.8639  &  0.0083  &  1.8652  &  0.0086  \\ 
3.29  &  1.8615  &  0.0097  &  1.8642  &  0.0092  \\ 
3.31  &  1.867  &  0.0087  &  1.8857  &  0.0092  \\ 
3.33  &  1.8777  &  0.0081  &  1.877  &  0.0085  \\ 
3.35  &  1.8692  &  0.0082  &  1.8629  &  0.0087  \\ 
3.37  &  1.8704  &  0.0088  &  1.8695  &  0.0088  \\ 
3.39  &  1.8844  &  0.0085  &  1.8799  &  0.0091  \\ 
3.41  &  1.8643  &  0.0084  &  1.8593  &  0.0085  \\ 
3.43  &  1.8699  &  0.0092  &  1.8643  &  0.0084  \\ 
3.45  &  1.8789  &  0.0089  &  1.8691  &  0.009  \\ 
3.47  &  1.8558  &  0.0101  &  1.8531  &  0.0094  \\ 
3.49  &  1.8654  &  0.0087  &  1.857  &  0.0085  \\ 
3.51  &  1.8549  &  0.0089  &  1.8583  &  0.0091  \\ 
3.53  &  1.8731  &  0.0093  &  1.8666  &  0.0092  \\ 
3.55  &  1.8637  &  0.0086  &  1.8658  &  0.0095  \\ 
3.57  &  1.8583  &  0.0095  &  1.863  &  0.0087  \\ 
3.59  &  1.8815  &  0.0096  &  1.8919  &  0.0097  \\ 
3.61  &  1.8552  &  0.009  &  1.8676  &  0.0101  \\ 
3.63  &  1.8673  &  0.0095  &  1.8881  &  0.0098  \\ 
3.65  &  1.8615  &  0.0105  &  1.8757  &  0.0098  \\ 
3.67  &  1.8592  &  0.0094  &  1.8744  &  0.0103  \\ 
3.69  &  1.8881  &  0.0109  &  1.8915  &  0.0104  \\ 
3.71  &  1.8779  &  0.0101  &  1.8915  &  0.0114  \\ 
3.73  &  1.8615  &  0.0093  &  1.8632  &  0.0101  \\ 
3.75  &  1.8415  &  0.0105  &  1.8485  &  0.0111  \\ 
3.77  &  1.8846  &  0.0111  &  1.8905  &  0.01  \\ 
3.79  &  1.8677  &  0.0112  &  1.8688  &  0.011  \\ 
3.81  &  1.8758  &  0.0099  &  1.8871  &  0.0108  \\ 
3.83  &  1.8753  &  0.0108  &  1.8757  &  0.0112  \\ 
3.85  &  1.8695  &  0.0104  &  1.8697  &  0.011  \\ 
3.87  &  1.8624  &  0.0103  &  1.8656  &  0.0111  \\ 
3.89  &  1.8613  &  0.0109  &  1.8715  &  0.012  \\ 
3.91  &  1.878  &  0.0116  &  1.8934  &  0.013  \\ 
3.93  &  1.8682  &  0.0115  &  1.8638  &  0.0119  \\ 
3.95  &  1.8588  &  0.0117  &  1.8721  &  0.0122  \\ 
3.97  &  1.8623  &  0.0109  &  1.8691  &  0.0122  \\ 
3.99  &  1.8828  &  0.0113  &  1.8705  &  0.0127  \\ 
    \hline
    \hline
    \end{tabular}
\hspace{24mm}
	\begin{tabular}{ccccc}
\\[0mm]
    \hline
	\hline \\[-3mm]
	   & \multicolumn{2}{c}{\teatro} & \multicolumn{2}{c}{\cascade} \\[0.5mm]
	  Wavelength   &  Depth  &  Uncertainty  &  Depth  &  Uncertainty \\
    $[\mu\rm m]$   & [\%] & [\%] & [\%] & [\%] \\[1mm]
	\hline \\[-3mm]
4.01  &  1.8754  &  0.0111  &  1.8651  &  0.0125  \\ 
4.03  &  1.8532  &  0.0111  &  1.8529  &  0.0141  \\ 
4.05  &  1.8599  &  0.0112  &  1.8831  &  0.0125  \\ 
4.07  &  1.863  &  0.0124  &  1.8753  &  0.0123  \\ 
4.09  &  1.8519  &  0.0123  &  1.8617  &  0.0126  \\ 
4.11  &  1.8801  &  0.0126  &  1.8908  &  0.0126  \\ 
4.13  &  1.8865  &  0.013  &  1.8982  &  0.0131  \\ 
4.15  &  1.8768  &  0.0121  &  1.8709  &  0.0133  \\ 
4.17  &  1.8729  &  0.0121  &  1.8811  &  0.0133  \\ 
4.19  &  1.885  &  0.0127  &  1.8914  &  0.014  \\ 
4.21  &  1.8954  &  0.0138  &  1.9104  &  0.0138  \\ 
4.23  &  1.9149  &  0.0123  &  1.908  &  0.0129  \\ 
4.25  &  1.908  &  0.0133  &  1.9122  &  0.0143  \\ 
4.27  &  1.9303  &  0.0131  &  1.9238  &  0.0134  \\ 
4.29  &  1.9265  &  0.0131  &  1.9316  &  0.0135  \\ 
4.31  &  1.9348  &  0.0136  &  1.925  &  0.0134  \\ 
4.33  &  1.9426  &  0.0132  &  1.9248  &  0.0133  \\ 
4.35  &  1.9602  &  0.0148  &  1.9498  &  0.0138  \\ 
4.37  &  1.9075  &  0.0145  &  1.9149  &  0.0138  \\ 
4.39  &  1.9121  &  0.0138  &  1.9117  &  0.014  \\ 
4.41  &  1.9151  &  0.0156  &  1.9053  &  0.0155  \\ 
4.43  &  1.9095  &  0.0141  &  1.9086  &  0.014  \\ 
4.45  &  1.884  &  0.0152  &  1.8829  &  0.0148  \\ 
4.47  &  1.8873  &  0.0153  &  1.9048  &  0.0146  \\ 
4.49  &  1.8984  &  0.0143  &  1.8813  &  0.0146  \\ 
4.51  &  1.8902  &  0.016  &  1.8904  &  0.0161  \\ 
4.53  &  1.8546  &  0.016  &  1.87  &  0.0157  \\ 
4.55  &  1.9002  &  0.0174  &  1.8872  &  0.016  \\ 
4.57  &  1.8956  &  0.0164  &  1.8865  &  0.016  \\ 
4.59  &  1.8785  &  0.0169  &  1.8839  &  0.0162  \\ 
4.61  &  1.8941  &  0.0172  &  1.8788  &  0.016  \\ 
4.63  &  1.9041  &  0.0187  &  1.8667  &  0.0167  \\ 
4.65  &  1.879  &  0.0175  &  1.9289  &  0.0175  \\ 
4.67  &  1.8866  &  0.0175  &  1.8898  &  0.017  \\ 
4.69  &  1.9101  &  0.0178  &  1.9194  &  0.0182  \\ 
4.71  &  1.9118  &  0.0186  &  1.9207  &  0.0183  \\ 
4.73  &  1.8845  &  0.0195  &  1.9253  &  0.018  \\ 
4.75  &  1.8628  &  0.0195  &  1.8823  &  0.0183  \\ 
4.77  &  1.9053  &  0.0196  &  1.8687  &  0.0196  \\ 
4.79  &  1.8754  &  0.0179  &  1.8853  &  0.0188  \\ 
4.81  &  1.8986  &  0.02  &  1.8768  &  0.0191  \\ 
4.83  &  1.9225  &  0.0185  &  1.92  &  0.0226  \\ 
4.85  &  1.9085  &  0.0201  &  1.9163  &  0.0227  \\ 
4.87  &  1.9154  &  0.0198  &  1.8781  &  0.0215  \\ 
4.89  &  1.8833  &  0.0215  &  1.882  &  0.022  \\ 
4.91  &  1.8996  &  0.0225  &  1.8951  &  0.0245  \\ 
4.93  &  1.91  &  0.0215  &  1.8934  &  0.0235  \\ 
4.95  &  1.9017  &  0.0238  &  1.8855  &  0.0239  \\ 
4.97  &  1.8978  &  0.0219  &  1.8699  &  0.0232  \\ 
4.99  &  1.8716  &  0.0219  &  1.8734  &  0.0225  \\ 
5.01  &  1.8561  &  0.0222  &  1.83  &  0.0243  \\ 
5.03  &  1.8872  &  0.0211  &  1.8733  &  0.0246  \\ 
5.05  &  1.8649  &  0.0207  &  1.8776  &  0.0237  \\ 
5.07  &  1.9117  &  0.0219  &  1.9019  &  0.0245  \\ 
5.09  &  1.864  &  0.0219  &  1.868  &  0.0249  \\ 
5.11  &  1.8776  &  0.0217  &  1.8707  &  0.0264  \\ 
5.13  &  1.8681  &  0.0222  &  1.8382  &  0.025  \\ 
5.15  &  1.856  &  0.0244  &  1.8749  &  0.025  \\ 
5.17  &  1.877  &  0.0276  &  1.8898  &  0.0414  \\     
    \hline
    \hline
	\end{tabular}
}
    \tablefoot{The bin size is 0.02~\mic and the wavelengths are the central wavelengths of each bin.}
\end{center}
\label{tab:spectrum}
\end{table}

\FloatBarrier

\newpage

\section{Retrieval posterior distributions from \hst \& NIRSpec}

\begin{figure*}[h]
    \includegraphics[width=0.95\textwidth]{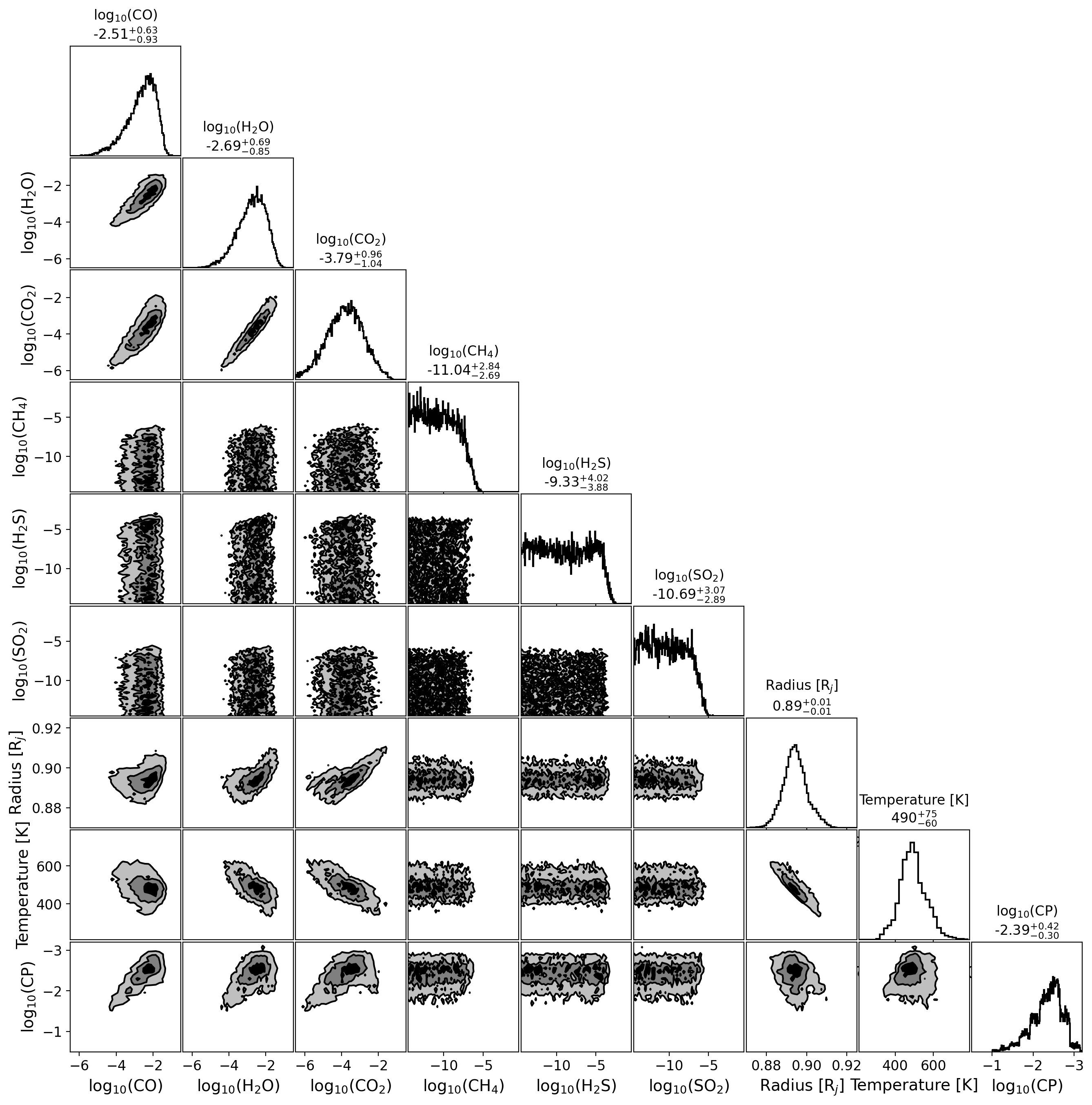}
    \caption{Posterior distributions for the \arcis retrieval on the combined \hst WFC3 G141 and \jwst NIRSpec G395M (\teatro) spectra. The molecular abundances are in volume mixing ratio, the planet radius is in R$_{\rm Jup}$ and is at 10~bar, the temperature is in Kelvin, and the cloud pressure is in bar. The steps in the cloud pressure posterior distribution reflect the atmospheric pressure sampling used in the retrievals.}
    \label{fig:posteriors}
\end{figure*}

\FloatBarrier

\newpage

\section{Reduced chi-square uncertainty}
\label{ap:chi-square}

We estimated the reduced chi-square uncertainty by\,$\sqrt{2/N}$, where the number of data points $N$ is 142. As mentioned in Sect.~3 of \citet{Andrae2010}, the reduced chi-square uncertainty can be approximated by\,$\sqrt{2/N}$ only for the true model having the true parameter values and when $N$ is large. In our fits, the true model and true parameters are unknown but the reduced chi-square shows that the best-fit model is a good fit to the data, and $N$ is large enough to approximate the chi-square distribution by a Gaussian, following Eq.~(14) and Fig.~2 of \citet{Andrae2010}. So we used\,$\sqrt{2/N}$ (i.e. 0.12) as a rough estimation of the reduced chi-square uncertainty, although it is not strictly valid.

\vspace{3mm}

\section{Abundances and detection significance}

\begin{table*}[h]
	\begin{center}
    \caption{Retrieved abundances and detection significance for all species considered here, obtained from the combined \hst WFC3 and \jwst \nirspec spectra.}
\renewcommand{\arraystretch}{1.2}
    \begin{tabular}{cccccccccc} \hline \hline \\[-3mm]
        & \multicolumn{3}{c}{\hst \& \teatro (\arcis)} &  \multicolumn{3}{c}{\hst \& \cascade (\arcis)} &  \multicolumn{3}{c}{\hst \& \teatro (\taurex)}\\[0.5mm]
     Molecule & log$_{10}$(VMR) & $\Delta$ln(E) & Sigma &  log$_{10}$(VMR) & $\Delta$ln(E) & Sigma & log$_{10}$(VMR) & $\Delta$ln(E) & Sigma \\[1mm] \hline \\[-3mm]
    CO$_2$& -3.79$^{+0.96}_{-1.04}$ &  71.3 & 12.2 & -3.05$^{+1.47}_{-2.11}$ & 51.4  & 10.4 & -3.39$^{+1.00}_{-1.21}$ & 69.8 & 12.0 \\
    CO & -2.51$^{+0.63}_{-0.93}$  &  6.7 & 4.1 & -3.19$^{+0.97}_{-2.19}$ & 2.5 & 2.8 & -2.11$^{+0.53}_{-0.99}$ & 5.4 & 3.7 \\
    H$_2$O& -2.69$^{+0.69}_{-0.85}$  &  16.0 & 6.0 & -2.09$^{+0.58}_{-1.61}$ & 13.0 & 5.4 & -2.46$^{+0.57}_{-0.91}$ & 15.4 & 5.9 \\
    H$_2$S&  $<-2.94$  &  0.2 & - &   -2.98$^{+0.59}_{-1.33}$  & 3.6  &  3.2 & $<-2.96$ & 0.3 & -  \\
    SO$_2$&  $<-5.33$  &  -0.2 & - &   $<-5.35$  & -0.4  & -  & $<-5.38$ & -0.2 & -  \\
    CH$_4$&  $<-5.73$  &  -0.2 & - &   $<-5.55$  & -0.3  & -  & $<-5.85$ & -0.3 & -  \\ 
    \\
     & log$_{10}$(P) [bar] & $\Delta$ln(E) & Sigma &  log$_{10}$(P) [bar] & $\Delta$ln(E) & Sigma & log$_{10}$(P) [bar] & $\Delta$ln(E) & Sigma \\[1mm] \hline \\[-3mm]
    Clouds & -2.39$^{+0.42}_{-0.30}$ & 8.8 & 4.6 & -1.71$^{+1.14}_{-0.58}$ & 1.5 & 2.3 & -2.32$^{+0.38}_{-0.29}$ & 8.4 & 4.5  \\[1mm] \hline \hline
    \end{tabular}
    \tablefoot{The retrieved cloud pressure is also given. The uncertainties are the $1\,\sigma$ uncertainties. Where the abundance was not constrained, the $3\,\sigma$ upper limit is given.}
	\end{center}
    \label{tab:retrievals}
\end{table*}

\FloatBarrier

\newpage

\section{Retrieval posterior distributions from NIRSpec only}

\begin{figure*}[h]
    \includegraphics[width=0.95\textwidth]{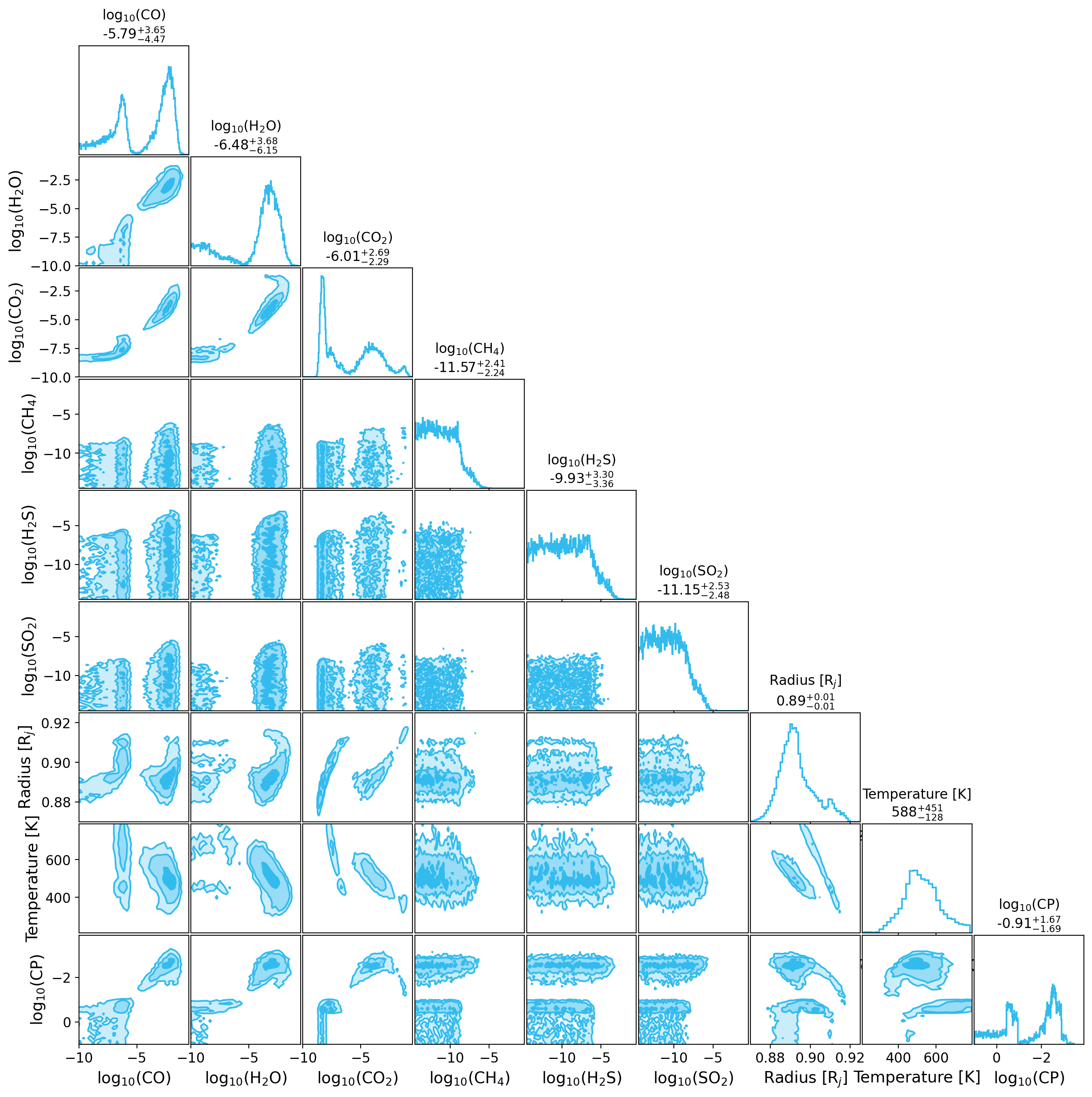}
    \caption{Posterior distributions for the \arcis retrieval on only the \jwst NIRSpec G395M (\teatro) spectrum. The molecular abundances are in volume mixing ratio, the planet radius is in R$_{\rm Jup}$ and is at 10~bar, the temperature is in Kelvin, and the cloud pressure is in bar.}
    \label{fig:posteriors_nirspec_only}
\end{figure*}

\FloatBarrier

\newpage

\section{Contribution of atmospheric regions to the transit spectrum}

\begin{figure*}[h]
    \centering
    \includegraphics[width=0.62\textwidth]{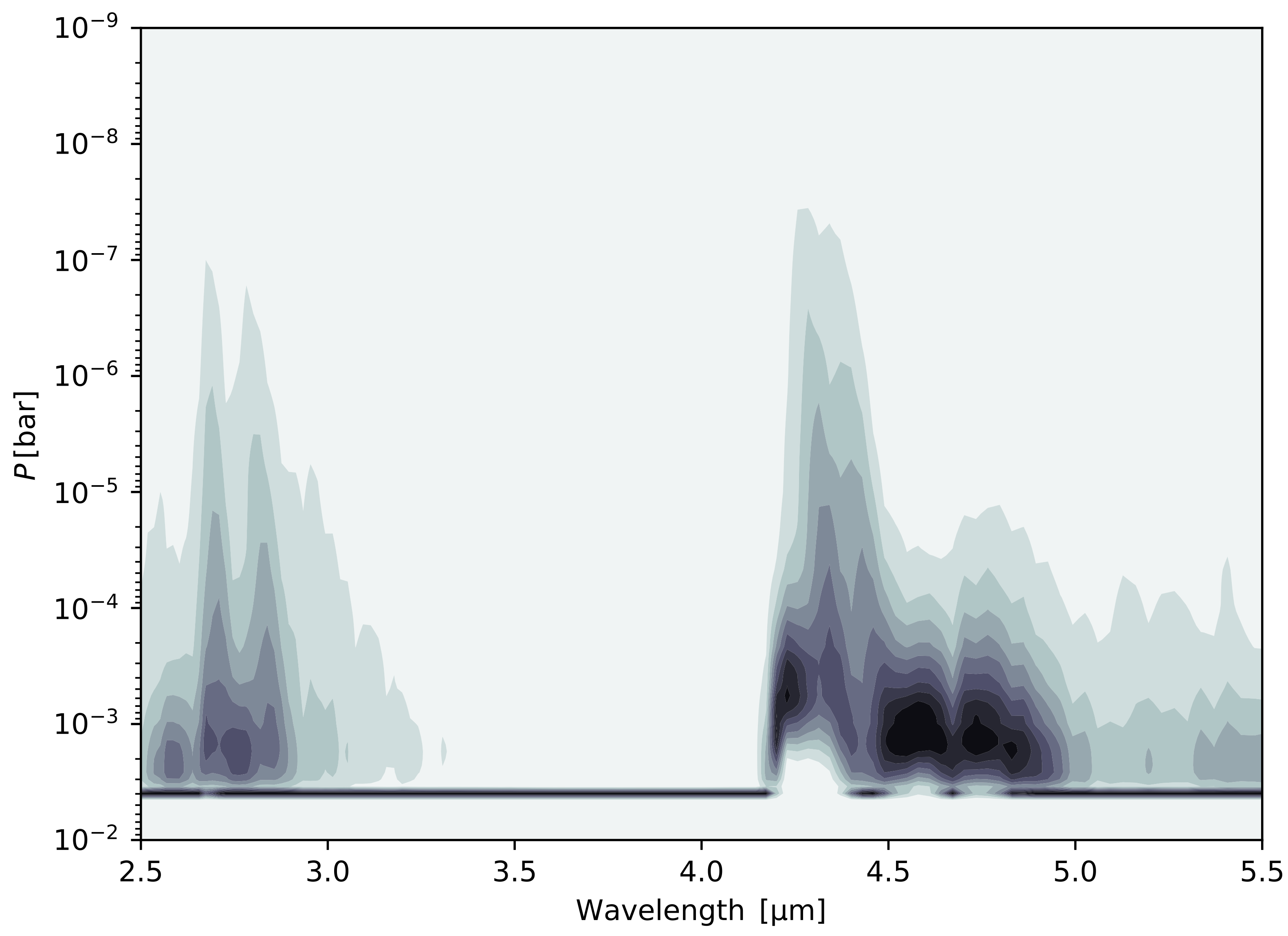}
    \caption{Contribution function of the median probability model that results from the \arcis retrieval on the combined \hst \& \nirspec (\teatro) spectra.}
    \label{fig:contribution}
\end{figure*}

The two-dimensional contribution function in Fig.~\ref{fig:contribution} shows which regions of the atmosphere contribute to the observed transit spectrum. These contributions are obtained by computing the difference in the transit spectrum created by clearing a single pressure layer in the model from all opacity sources. At each wavelength, it shows at which heights in the atmosphere the transit spectrum is created. At the spectral location of the CO$_2$ feature, the transit spectrum is influenced by the upper layers of the atmosphere, while a large part of the \nirspec spectrum is mostly created at the cloud top level.

\section{Temperature dependence of \ce{SO2} production}

\begin{figure*}[h]
    \centering
    \includegraphics[width=0.64\columnwidth]{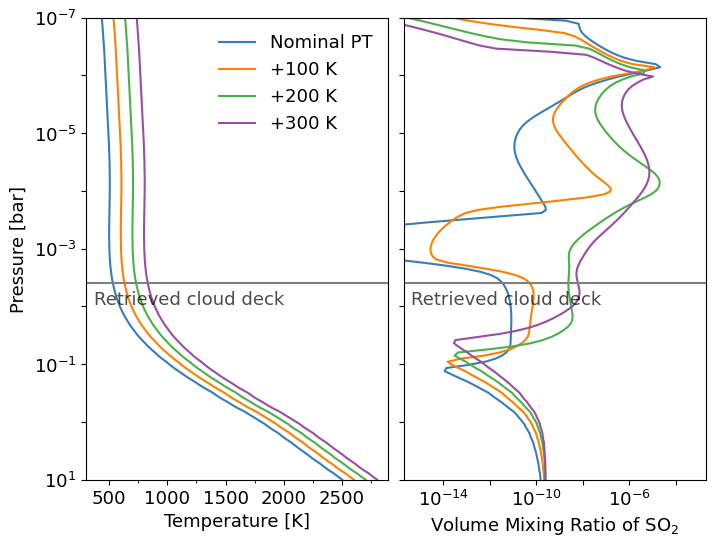}
    \caption{Volume mixing ratios of \ce{SO2} that result from \vulcan simulations with different $P-T$ structures for \mbox{HAT-P-12\,b}. The nominal model corresponds to the $P-T$ profile used throughout this study, a $10\times$ solar metallicity, a C/O ratio of 0.50, with photochemistry. For each subsequent model, the temperature is increased by steps of 100~K; the other parameters are unchanged.}
    \label{fig:SO2temperature}
\end{figure*}

\end{appendix}

\end{document}